\documentclass[12pt]{article}

\usepackage{fullpage}
\usepackage{cite}
\usepackage{graphicx}
\usepackage{amsmath}
\usepackage{amssymb}
\usepackage{color}
\usepackage{graphics}
\usepackage{xcolor,colortbl}
\definecolor{grey}{rgb}{0.8, 0.8, 0.8}
\definecolor{lightgrey}{rgb}{0.94, 0.94, 0.94}
\definecolor{darkgreen}{rgb}{0.00, 0.50, 0.00}

\usepackage{hyperref}

\allowdisplaybreaks
\title{}
\date{}

\def\beq{\begin{equation}}
\def\eeq{\end{equation}}
\def\beqa{\begin{eqnarray}}
\def\eeqa{\end{eqnarray}}
\def\eq#1{Eq.~(\ref{#1})}
\renewcommand{\vec}[1]{\mbox{\boldmath$ #1 $}}
\newcommand{\e}{\epsilon}
\newcommand{\secn}[1]{Section~\ref{#1}}

\newcommand{\ee}{\eeq}
\newcommand{\bea}{\begin{eqnarray}}
\newcommand{\eea}{\end{eqnarray}}
\newcommand{\nn}{\nonumber}

\newcommand{\as}{\alpha_s}
\newcommand{\eps}{\epsilon}

\newcommand \slsh [1] {\not\!{#1}}

\begin{document}


\bibliographystyle{utphys}

\titlepage

\begin{flushright}
QMUL-PH-19-12\\
Nikhef/2019-015\\
MS-TP-19-08
\end{flushright}

\vspace*{1.2cm}

\begin{center}
{\Large \bf Diagrammatic resummation of leading-logarithmic threshold effects at
  next-to-leading power}

\vspace*{1cm} \textsc{N. Bahjat-Abbas$^a$,
  D. Bonocore$^b$,
  J. Sinninghe Damst\'{e}$^{c,d}$,
  E. Laenen$^{c,d,e}$,\\
  L. Magnea$^{f}$,
  L. Vernazza$^{d}$ and
  C. D. White$^a$} \\

\vspace*{1.2cm} 

$^a$ Centre for Research in String Theory, School of Physics and
Astronomy, Queen Mary University of London, 327 Mile End Road, London
E1 4NS, UK

\vspace*{0.2cm}

$^b$ Institut f\"{u}r Theoretische Physik, Westf\"{a}lische
Wilhelms-Universit\"{a}t M\"{u}nster, Wilhelm-Klemm-Stra\ss e 9,
D-48149 M\"{u}nster, Germany\\

\vspace*{0.2cm}

$^c$ ITFA, University of Amsterdam, Science Park 904, Amsterdam, 
The Netherlands \\

\vspace*{0.2cm} 

$^d$ Nikhef, Science Park 105, NL-1098 XG Amsterdam, The Netherlands

\vspace*{0.2cm} 

$^e$ ITF, Utrecht University, Leuvenlaan 4, Utrecht, The Netherlands

\vspace*{0.2cm} 

$^f$ Dipartimento di Fisica Teorica and Arnold-Regge Center, 
Universit\`a di Torino, and \\
INFN, Sezione di Torino, Via Pietro Giuria 1, I-10125 Torino, Italy

\end{center}

\vspace*{7mm}

\begin{abstract}
  \noindent Perturbative cross-sections in QCD are beset by logarithms
  of kinematic invariants, whose arguments vanish when heavy particles
  are produced near threshold. Contributions of this type often need to be 
  summed to all orders in the coupling, in order to improve the behaviour 
  of the perturbative expansion, and it has long been known how to do 
  this at leading power in the threshold variable, using a variety of approaches. 
  Recently, the problem of extending this resummation to logarithms suppressed 
  by a single power of the threshold variable has received considerable attention. 
  In this paper, we show that such next-to-leading power (NLP) contributions 
  can indeed be resummed, to leading logarithmic (LL) accuracy, for any 
  QCD process with a colour-singlet final state, using a direct generalisation 
  of the diagrammatic methods available at leading power. We compare our 
  results with other approaches, and comment on the implications for further 
  generalisations beyond leading-logarithmic accuracy.
\end{abstract}


\section{Introduction}
\label{intro}

Perturbative calculations of hadronic cross sections in Quantum 
Chromodynamics (QCD) are the cornerstone of theoretical predictions 
for all processes of phenomenological interest at particle colliders, 
such as the Large Hadron Collider (LHC). Furthermore, the ever-increasing
precision of experimental data demands that theoretical predictions 
for scattering processes of interest be continually improved. The 
relevant calculations are carried out using an expansion in powers 
of the coupling constant $\alpha_s$, and typically proceed on two 
fronts. First, one may determine the complete behaviour of a given 
quantity at a fixed order in the coupling constant. The state of the art 
for most processes is next-to-leading order (NLO) in perturbation theory, 
with an increasing number of notable exceptions known at NNLO, and 
even N$^3$LO (see e.g.~\cite{Tanabashi:2018oca} for a review). 
Whilst successful for many observables, the fixed order approach is 
only valid provided subleading perturbative corrections are well-behaved. 
Given that perturbative coefficients depend on the momenta of the 
scattering particles, this criterion can fail in certain kinematic regimes: 
a well-known example is the production of heavy particles near threshold. 
In such processes, one can define a {\it threshold variable} $\xi$, which 
satisfies $\xi\rightarrow 0$ when the heavy particles carry all of the energy 
in the final state. The precise definition of $\xi$ will depend on the process 
being considered, but, generically, it has the form of a dimensionless ratio 
of kinematic invariants. One may then write a general schematic form for
production cross-sections near threshold, as
\beq
  \frac{d \sigma}{d \xi} \, = \, \sigma_0 \sum_{n = 0}^{\infty} 
  \left( \frac{\alpha_s}{\pi} \right)^n \sum_{m = 0}^{2 n - 1} 
  \left[ \, c_{n m}^{(-1)} \left( \frac{\log^m \xi}{\xi} \right)_+ 
  + \, c_{n}^{(\delta)} \, \delta(\xi) + \, c_{nm}^{(0)} \, \log^m \xi + 
  {\cal O}(\xi) \, \right] \, .
\label{thresholddef}
\eeq
Here we denote by $\sigma_0$ the Born-level cross section, which may 
contain additional coupling factors. The first contribution in the square 
brackets consists of a series of terms, at fixed order in $\alpha_s$, 
containing powers of the logarithm of the threshold variable, divided by 
$\xi$ itself. These contributions can be directly traced to soft and 
collinear singularities of the underlying scattering amplitudes: the 
cancellation of infrared divergences between virtual corrections and 
real radiation leaves behind potentially large corrections, which are 
still singular as $\xi \rightarrow 0$, but are regularised by the well-known 
plus prescription, so that they are integrable; as discussed below, the 
all-order structure of these terms is well understood. The second set 
of terms in \eq{thresholddef} has support localised on the threshold,
$\xi = 0$, and for processes with electroweak final states it is known
that such terms can be formally exponentiated (see, for example, 
Ref.~\cite{Eynck:2003fn}). The third set of terms in the square brackets 
is suppressed by a single power of $\xi$ with respect to the leading-power
contribution. Although formally not as divergent as the preceding terms,
they are nevertheless still singular as $\xi \rightarrow 0$, and thus
potentially numerically sizeable in the threshold region. These next-to-leading
power (NLP) terms are the focus of the present work, while we will
neglect all further subleading contributions to \eq{thresholddef}, which
vanish at threshold.

Order by order in perturbation theory, one can distinguish two expansions 
in \eq{thresholddef}. Firstly, there is an expansion in powers of the threshold 
variable $\xi$, in which we can distinguish the plus distributions and delta 
function terms as being {\it leading power} (LP) in $\xi$, while the remaining 
logarithms are {\it next-to-leading-power} (NLP). Secondly, for each fixed
power of $\xi$, we can consider the expansion in powers of the logarithm, 
labelling terms proportional to $\log^{2 n - 1} \xi$ as {\it leading logarithmic} 
(LL), the next-highest power as {\it next-to-leading logarithmic} (NLL), and 
so on. The problematic nature of LP terms was noted already in the early 
days of QCD (see for example~\cite{Altarelli:1979ub}), and it was quickly 
realised that such terms at LL level could be summed up to all orders in 
perturbation theory to achieve a well-behaved result as $\xi \rightarrow
0$~\cite{Parisi:1980xd,Curci:1979am}. This resummation was subsequently 
extended to subleading logarithmic accuracy using two equivalent
approaches~\cite{Sterman:1987aj,Catani:1989ne,Catani:1990rp},
themselves partially reliant on earlier diagrammatic arguments for the
exponentiation of soft behaviour~\cite{Gatheral:1983cz,Frenkel:1984pz,
Sterman:1981jc}. Since that time, LP threshold resummation has been 
reinterpreted and clarified using a wide variety of methods, including the 
use of Wilson lines~\cite{Korchemsky:1993xv,Korchemsky:1993uz}, the 
renormalisation group~\cite{Forte:2002ni}, the connection to factorisation
theorems~\cite{Contopanagos:1997nh}, and soft collinear effective
theory (SCET)~\cite{Becher:2006nr,Schwartz:2007ib,Bauer:2008dt,
Chiu:2009mg}. The state of the art for resummation at LP is NNLL accuracy 
in many processes, including some cases of differential distributions. Recent, 
pedagogical reviews may be found in Refs.~\cite{Luisoni:2015xha,
Becher:2014oda,Campbell:2017hsr}.

The phenomenological success of LP resummation, together with the 
increasing precision of contemporary collider data, makes it natural to 
ponder whether NLP terms in the threshold expansion can also be classified
and resummed, particularly since they have been shown to be numerically 
significant in important scattering processes~\cite{Kramer:1996iq,
Herzog:2014wja}. Indeed, the study of such contributions has a long 
history. Subleading corrections involving soft momenta were first investigated 
in the classic works of Refs.~\cite{Low:1958sn,Burnett:1967km}, which 
dealt exclusively with massive particles in QED. The analysis of 
Ref.~\cite{DelDuca:1990gz} updated this to include massless particles. 
Some years later, the topic was investigated using path-integral 
methods in Ref.~\cite{Laenen:2008gt}, which derived a set of effective 
Feynman rules for the emission of gauge bosons at next-to-soft level, and
argued that a large class of NLP contributions exponentiates. The results 
were subsequently confirmed by an all-order analysis of Feynman 
diagrams~\cite{Laenen:2010uz}, but concerned massive partons only. 
In a different approach, NLP effects in certain processes were argued to be 
resummable, based on well-motivated physical assumptions~\cite{Soar:2009yh,
Moch:2009hr,Moch:2009mu,deFlorian:2014vta,Presti:2014lqa} (see also
Refs.~\cite{Akhoury:1998gs,Laenen:2008ux,Grunberg:2007nc,Grunberg:2009yi,
Grunberg:2009vs} for other work related to elucidating all-order properties).

More recently, there has been a revival of interest in studying NLP effects at 
amplitude level, partly motivated by more formal work relating soft radiation 
to asymptotic symmetries of the $S$-matrix in gauge theories and in
gravity~\cite{Cachazo:2014fwa,Casali:2014xpa}. Thus, in addition to the 
phenomenological applications mentioned above, the study of subleading 
threshold effects in quantum field theory can have a role to play in finding 
new representations of, and relations between, gauge and gravity
theories~\cite{Naculich:2011ry,White:2011yy,Oxburgh:2012zr,Akhoury:2011kq,
Laddha:2018myi,Sahoo:2018lxl}, whilst also finding applications in transplanckian 
scattering~\cite{Saotome:2012vy, Akhoury:2013yua,Melville:2013qca,
Luna:2016idw}. In the latter context (as potentially in gauge theories), 
resummation plays a key role.

In QCD (and related gauge theories), threshold resummation at leading
power is known to be a consequence of the universal factorisation of
soft and collinear divergences in scattering amplitudes (see for
example Ref.~\cite{Contopanagos:1997nh} for a dedicated discussion of
this point).  This has motivated attempts to construct a factorisation
formula for NLP
effects. References~\cite{Bonocore:2015esa,Bonocore:2016awd} use a
diagrammatic approach, building on the earlier work of
Ref.~\cite{DelDuca:1990gz}, to describe the effect of dressing a
general non-radiative amplitude with an additional gluon emission up
to NLO, and NLP in the threshold expansion.  This formula contains
universal functions similar to those appearing at LP level, but
including extra contributions that describe, for example, the emission
of wide-angle soft gluons from within jets. A more complete analysis
for scalar theories coupled to electromagnetism was undertaken in
Refs.~\cite{Gervais:2017yxv, Gervais:2017zky,Gervais:2017zdb}, which
again stress the importance of new quantities (both universal and
non-universal) that appear beyond LP order in emitted gluon
momentum. Related analyses have been carried out in
SCET~\cite{Larkoski:2014bxa,Kolodrubetz:2016uim,Moult:2016fqy,Moult:2017rpl,
  Feige:2017zci,Chang:2017atu} (see Ref.~\cite{Beneke:2004in} for
earlier work in the context of flavour physics), and results using
either diagrammatic or effective theory methods have been shown to be
potentially useful for improving the accuracy of fixed-order
calculations~\cite{DelDuca:2017twk,Bonocore:2014wua,
  Bahjat-Abbas:2018hpv,Boughezal:2016zws,Boughezal:2018mvf,Moult:2016fqy,Moult:2017jsg,Ebert:2018lzn,Ebert:2018gsn,vanBeekveld:2019cks,vanBeekveld:2019prq}. Recently,
the SCET framework has been used to demonstrate that the
leading-logarithmic (LL) NLP contributions can be resummed, first for
event shapes~\cite{Moult:2018jjd}, and then for Drell-Yan
production~\cite{Beneke:2018gvs}, where the results agree with the
predictions of the {\it physical evolution kernel} approach of
Refs.~\cite{Soar:2009yh,Moch:2009hr,Moch:2009mu,deFlorian:2014vta,
  Presti:2014lqa}.

Our aim in this paper is to show how a similar resummation of LL NLP
effects can be achieved using the diagrammatic approach developed in
Refs.~\cite{Laenen:2008gt,Laenen:2010uz}, and itself analogous to the
original LP resummations of Refs.~\cite{Sterman:1987aj,Catani:1989ne,
  Gatheral:1983cz,Frenkel:1984pz,Sterman:1981jc}. As in the SCET
approach of Ref.~\cite{Beneke:2018gvs} (and as observed in
Refs.~\cite{Bonocore:2015esa,
  Bonocore:2016awd,Gervais:2017yxv,Gervais:2017zky,Gervais:2017zdb}),
we will see that, while it is true that a number of new functions
appear at NLP level in the threshold expansion, many of them are
irrelevant for discussing the highest power of the NLP logarithm at
any given order in perturbation theory. Thus, the resummation of LL
NLP contributions is remarkably straightforward. Importantly, this
method is sufficiently simple and universal that it can be directly
applied to any hadronic cross section with colour-singlet final
states: indeed, we explicitly discuss applications to Higgs boson
production in the gluon fusion channel, and the formalism can readily
be generalised to multi-boson final states. There are a number of
motivations for the present analysis. First, they are a natural
application of the programme of work commenced in
refs.~\cite{Laenen:2008gt, Laenen:2010uz}, where it is was shown that
a broad subclass of NLP effects indeed exponentiates. Second, the
history of LP resummation suggests that it is highly useful to have
more than one formalism for describing equivalent physics: comparison
of one approach with another has the potential to clarify both, and it
may also be the case that different approaches have relative strengths
and weaknesses, thus being more or less suitable in any given
context. Third, our diagrammatic approach will provide an alternative
starting point for generalising the NLP resummation formalism beyond
leading-logarithmic accuracy.

The structure of our paper is as follows. In \secn{sec:LP}, we review the 
resummation of LP threshold contributions, introducing notation that will 
be useful for what follows. In particular, we will relate our calculation to the 
path-integral methods of Ref.~\cite{Laenen:2008gt}, which provide a particularly 
elegant proof of exponentiation. In \secn{sec:NLP}, we show how the picture
can be naturally extended to NLP level, using existing results. We will argue 
in detail that potential additional contributions to NLP behaviour, including 
hard collinear effects, non-universal behaviour and phase-space correlations 
between gluons, can be ignored at LL. Armed with this knowledge, we will 
then perform an explicit calculation that resums the LL NLP terms in
Drell-Yan, comparing our results with others in the literature~\cite{Soar:2009yh,
Moch:2009hr,Moch:2009mu,deFlorian:2014vta,Presti:2014lqa}. We
will then comment on the general applicability of our framework to 
the production of an arbitrary number of colour singlet particles, before 
examining Higgs production in the large top mass limit as a further example. 
In \secn{sec:SCET}, we briefly compare our framework with the recent analysis 
of Ref.~\cite{Beneke:2018gvs}, in the framework of Soft Collinear Effective 
Theory (SCET). Finally, we discuss our results in \secn{sec:discuss} before
concluding. Technical details are collected in three appendices.


\section{Threshold resummation at leading power}
\label{sec:LP}

\begin{figure}
\begin{center}
  \scalebox{0.5}{\includegraphics{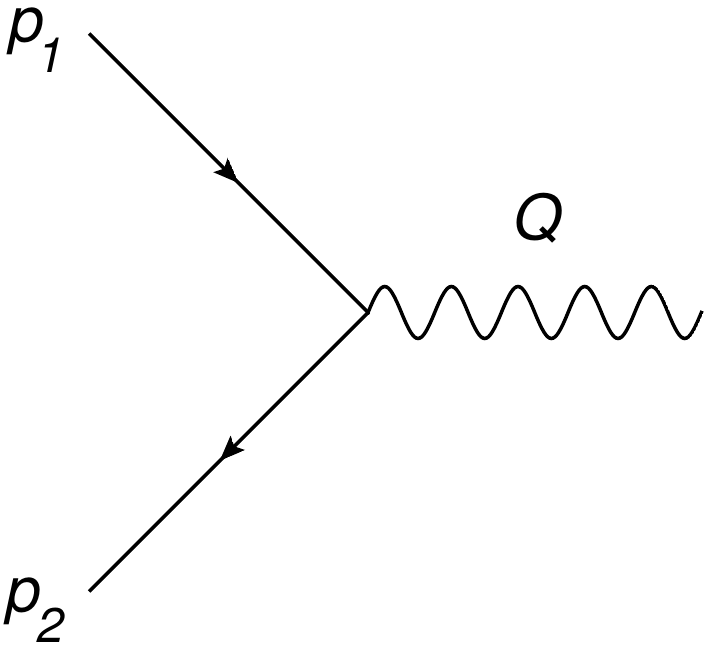}}
  \caption{Drell-Yan production at leading order.}
\label{fig:DY}
\end{center}
\end{figure}

In this section, we review the resummation of terms at leading power
in the threshold variable, using factorisation methods. Given that our aim 
in what follows is to sum leading-logarithmic terms only at NLP, we will 
mostly concern ourselves here with LL terms also at LP. Furthermore, 
we will phrase our discussion in terms of methods and notation that allow 
a straightforward generalisation to subleading power in the threshold 
expansion. While our discussion applies to general colour-singlet 
final states, we will first explicitly consider the Drell-Yan production of a 
massive (or off-shell) vector boson, which at LO corresponds to the 
partonic scattering process
\beq
  q(p_1) + \bar{q}(p_2) \, \rightarrow \, V(Q) \, ,
\label{DYLOproc}
\eeq
depicted in Figure~\ref{fig:DY}. We will not explicitly consider here the 
quark-gluon production channel, where NLP logarithms are present, but 
constitute in fact the leading power, since LP logarithms are absent; in 
the gluon-gluon channel for the Drell-Yan process, only NNLP logarithms 
can arise.
We write the invariant mass distribution 
in the $q\bar q$ channel as
\beq 
  \frac{d \sigma}{d \tau} \, = \, \sigma_0(Q^2) \int _0^1 d z d x_1
  d x_2 \, \delta ( \tau - x_1 x_2 z ) \, q (x_1, \mu_F^2) \, 
  \bar{q} (x_2, \mu_F^2) \,
  \Delta \! \left( z, \alpha_s(\mu_R^2),
  \frac{\mu_F^2}{Q^2}, \frac{\mu_R^2}{Q^2} \right) \, ,
\label{sigmahad}
\eeq
where we restrict ourselves to a single quark flavour for simplicity.
Here $\sigma_0 (Q^2)$ is the total cross section, whose precise 
value will depend on the nature of the vector boson. Furthermore, 
$\alpha_s(\mu_R^2)$ is the strong coupling at the renormalisation 
scale $\mu_R$, $q (x, \mu_F^2)$ is a quark distribution function with 
longitudinal momentum fraction $x$ and factorisation scale $\mu_F$, 
while $\bar{q}$ is the equivalent for an antiquark. Given that scale 
choice effects contribute to only subleading logarithms (see for 
example~\cite{Catani:1989ne}), we will simply choose $\mu_F 
= \mu_R = Q$ from now on, and simplify notation accordingly.
In \eq{sigmahad} we defined the invariants 
\beq
  \tau \, = \, \frac{Q^2}{s} \, , \qquad \qquad  z \, = \, \frac{Q^2}{\hat{s}} \, , 
\label{tau-zdef}
\eeq
where $\hat{s} = ( p_1 + p_2 )^2$ is the squared partonic centre of mass 
energy, and $s = \hat{s}/x_1 x_2$ is the hadronic centre of mass energy.
The ratio $z$ represents the fraction of $\hat{s}$ carried by the final
state vector boson. At LO this must be unity, so that one has 
\beq
  \Delta^{(0)} \left( z \right) \, = \delta(1 - z) \, .
\label{sigmaLO}
\eeq
The invariant mass distribution in \eq{sigmahad} is a convolution in $z$, 
and can be diagonalised by taking Mellin moments with respect to $\tau$, 
with the result
\beq
  \int_0^1 d \tau \, \tau^{N - 1} \, \frac{d \sigma}{d\tau} \, = \, 
   \sigma_0 (s,Q^2) \, q (N, Q^2) \, {\bar{q}} (N, Q^2) \, \Delta(N,Q^2) \, ,
\label{sigmahadN}
\eeq
where 
\beq
  q (N, Q^2) \, = \, \int dx \, x^{N - 1} \, q(x, Q^2)
\label{qN}
\eeq
is the transformed quark distribution (and similarly for the antiquark), 
and we have defined
\beq
  \Delta (N, Q^2) \, = \, \int_0^1 dz \, z^{N - 1} \, \Delta(z,Q^2).
\label{DeltaN}
\eeq
Beyond LO, \eq{DeltaN} receives potentially large threshold corrections. 
In Mellin space, these appear as contributions of the form
\beq
  \alpha_s^n \, \log^{m} N \, , \qquad m \, = \, 0, \ldots, 2 n \, , 
\label{Mellinlogs}
\eeq
which in momentum space are associated with plus distributions of the form
\beq
  {\cal D}_i \, = \, \left( \frac{\log^i(1 - z)}{1 - z} \right)_+ \, , 
  \qquad i \, = \, 0, \ldots 2 n - 1 \, ,
\label{plusdef}
\eeq
defined such that
\beq
  \int_0^1 dz \, f(z) \, \big[ g(z) \big]_+ \, = \, 
  \int_0^1 dz \big[ f(z) - f(1) \big] \, g(z) \, .
\label{plusdef2}
\eeq
When computed in perturbation theory from quark scattering, $\Delta 
\left(N,Q^2\right)$ is affected by collinear divergences, which must be 
reabsorbed in the quark distributions: below, we will mostly work with
the `bare' $\Delta$, before renormalisation of the coupling $\alpha_s$, 
and before the factorisation of collinear divergences, which will be
regulated using dimensional regularisation in $d = 4 -2 \epsilon$. For
clarity, we will denote this bare partonic cross section with 
$\widehat{\Delta}(z, Q^2, \epsilon)$ in momentum space, and with 
$\widehat{\Delta} (N, Q^2, \epsilon)$ in Mellin space. Collinear 
factorisation is understood to be performed in the 
$\overline{\rm MS}$ scheme.

For any QCD process with a colour-singlet final state produced near 
threshold, $\widehat{\Delta}$ has a factorised structure, and can be written 
as~\cite{Sterman:1987aj,Kidonakis:1997gm}
\beq
  \widehat{\Delta} \left( N, Q^2, \epsilon \right) \, = \, 
  \left| \mathcal{H} \left( Q^2 \right) \right|^2 \, 
    \frac{\prod_i \psi_i \left( N, Q^2, \epsilon \right)}
  {\prod_{i} \psi_{{\rm eik} ,i} \left( N, Q^2, \epsilon \right)} \, 
  {\cal S} \left( N, Q^2, \epsilon \right) \, ,
\label{Deltafac}
\eeq
where $\mathcal{H} \left(Q^2\right)$ is an amplitude-level finite \textit{hard 
function} containing off-shell virtual contributions, ${\cal S} (N, Q^2,
\epsilon)$ is a {\it soft function} collecting all soft enhancements associated 
with (real or virtual) soft radiation, and $\psi_i (N, Q^2, \epsilon)$ is a 
perturbative \textit{(anti-)quark distribution function}, collecting collinear 
singularities associated with initial line $i$; finally, given that infrared
enhancements of both soft and collinear origin are included twice (both 
in the soft and quark distribution functions), one may remove the double 
counting by dividing each quark distribution by its own eikonal approximation
$\psi_{ {\rm eik}, i} (N, Q^2, \epsilon)$. Formal definitions of the (eikonal)
quark distributions and of the soft function are given, for example, in 
Ref.~\cite{Kidonakis:1997gm}: sometimes, eikonal quark distributions 
are absorbed into the soft function to build the so-called {\it reduced soft 
function}, organising wide-angle soft radiation. On the other hand, one 
may consider the factor
\beq
  \psi_{\,{\rm h}, i} (N, Q^2, \epsilon) \, = \, \frac{\psi_i (N, Q^2, \epsilon)}
  {\psi_{{\rm eik}, i} (N, Q^2, \epsilon)} \, ,
\label{Jbardef}
\eeq
for each initial parton line: this has the effect of removing the soft behaviour 
from each quark distribution, leaving hard collinear behaviour only. This 
arrangement is particularly convenient if one wishes to focus only on leading 
logarithms, as we do in this paper: indeed, at any fixed order in $\alpha_s$,
leading logarithms at leading power arise only when the maximum number 
of singular integrations is performed, yielding the highest inverse power of 
$\epsilon$. Thus, the factor $\psi_{{\rm h}, i} (N, Q^2, \epsilon)$ for each 
external line contributes only at subleading logarithmic accuracy, and 
can be put equal to unity at LL.  We are then left with the simple result
\beq
  \widehat{\Delta} \left( N, Q^2, \epsilon \right) \, = \, 
  \left| \mathcal{H} \left( Q^2 \right) \right|^2
  {\cal S} \! \left( N, Q^2, \epsilon \right) \, ,
\label{sigmahadN2}
\eeq
implying that leading logarithms in the DY cross-section at arbitrary 
orders in perturbation theory are governed purely by the soft 
function~\cite{Sterman:1987aj,Catani:1989ne,Korchemsky:1993uz}, 
on which we now focus. 

For any QCD process with a colour-singlet final state, the soft function 
has a formal definition as a vacuum expectation value of Wilson line
operators associated with the colliding partons. Defining the dimensionless 
four-vectors $\beta_i$ via
\beq
  p_i^\mu \, = \, \sqrt{\hat{s}} \, \beta_i^\mu \, ,
\label{ndef}
\eeq
one may write the soft function (in momentum space) as
\beq
  {\cal S} (z, Q^2, \epsilon) \, = \, \frac{1}{N_c} \sum_n \,
  {\rm Tr} \Big[ \left\langle 0 \left|
  \Phi_{\beta_1}^\dag \Phi_{\beta_2} \right| n \right\rangle
  \left\langle n \left| \Phi_{\beta_2}^\dag \Phi_{\beta_1} \right| 0 \right\rangle
  \Big] \, \delta \!\left( z - \frac{Q^2}{\hat{s}} \right) \, . 
\label{Sdef}
\eeq
Here the trace is over colour indices, and the Wilson line operators are 
defined by
\beq
  \Phi_{\beta_i} \, = \, {\cal P} \exp \left[ {\rm i} g_s {\bf T}^a  \int_{-\infty}^0
  d \lambda \, \beta_i \cdot A ( \lambda \beta_i) \right] \, ,
\label{Phidef}
\eeq
where ${\bf T}^a$ is a colour generator in the fundamental 
representation; furthermore, \eq{Sdef} includes
a sum over final states containing $n$ partons generated by the Wilson 
lines, including the appropriate phase space integration, and subject to 
the constraint that the total energy radiated in the final state equal $(1 - z) 
\hat{s}$; finally, the division by the number of colours $N_c$ corrects for 
the fact that this factor has already been included in the LO cross-section
$\sigma_0$ in \eq{sigmahadN2}. Introducing the momentum space
gauge field $\tilde{A}_\mu(k)$, one may write the Wilson line exponent
as
\beq
  {\rm i} g_s {\bf T}_a \int \frac{d^d k}{(2\pi)^d} \, \beta_i \cdot \tilde{A}(k)
  \int_{- \infty}^0 d \lambda \, e^{{\rm i} \lambda  \beta_i \cdot k} 
  \, = \, \int \frac{d^d k}{(2\pi)^d} \, \tilde{A}_\mu(k) \left[ g_s {\bf T}^a \,
  \frac{\beta_i^\mu}{\beta_i \cdot k - {\rm i} \varepsilon} \right] \, ,
\label{Wilsonexp}
\eeq
where the square-bracketed factor on the right constitutes the
momentum-space factor associated to the emission of a gluon from the
Wilson line. We recognise this as the well-known {\it eikonal Feynman 
rule} for soft gluon emission, so that finding the soft function amounts 
to calculating the cross section for the incoming partons in the eikonal 
approximation. This cross-section is known to exponentiate, which
relies on two properties: first, vacuum expectation values of Wilson
lines exponentiate before any phase space integrations are carried
out, which may be shown diagrammatically~\cite{Gatheral:1983cz,
Frenkel:1984pz,Sterman:1981jc}, or using renormalisation group 
arguments, themselves relying on the multiplicative renormalisability 
of Wilson line operators~\cite{Polyakov:1980ca,Arefeva:1980zd,
Dotsenko:1979wb,Brandt:1981kf,Korchemsky:1985xj,Korchemsky:1987wg}; 
second, the phase space for the emission of $n$ soft partons factorises
into $n$ decoupled one-parton phase space integrals, given that 
momentum conservation can be ignored at leading power in the 
threshold expansion.

Combining these two properties, one finds that the complete soft
function, at cross-section level, has an exponential form, and the
exponent can be directly computed in terms of a special class of 
Feynman diagrams known as {\it webs}~\cite{Gatheral:1983cz,
Frenkel:1984pz,Sterman:1981jc}. These results have been reinterpreted 
more recently using a path integral approach~\cite{Laenen:2008gt}, 
which incorporated statistical physics methods (the {\it replica trick}) 
to provide a particularly streamlined proof of diagrammatic exponentiation. 
These methods have in turn allowed the web language to be generalised 
to multiparton scattering~\cite{Gardi:2010rn,Gardi:2011yz,Gardi:2013ita,
Gardi:2011wa,Dukes:2013wa,Dukes:2013gea,Gardi:2013saa,Falcioni:2014pka,
Dukes:2016ger} (see also~\cite{Mitov:2010rp,Vladimirov:2015fea}, or
Ref.~\cite{White:2015wha} for a pedagogical review). We review the
replica trick here in Appendix~\ref{app:replica}, given that it can also be 
used to demonstrate directly the exponentiation of a large class of
contributions at NLP in the threshold expansion.

\begin{figure}
\begin{center}
  \scalebox{1.0}{\includegraphics{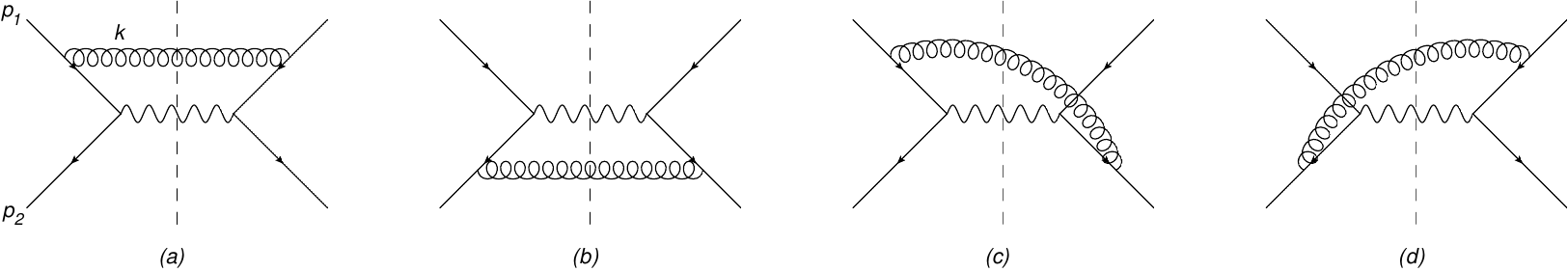}}
  \caption{Real emission diagrams for the eikonal cross-section of
  \eq{eikxsec}, where all emission vertices are assumed to be eikonal.}
\label{fig:NLOdiags}
\end{center}
\end{figure}

Concentrating on leading logarithms, it is important to note that the pattern
of exponentiation of soft and collinear singularities is non-trivial, in that the
exponent is single-logarithmic (containing terms of the form $\alpha_s^n 
\log^m N$ with $m \leq n+1$), while the cross section is double-logarithmic,
as noted in \eq{Mellinlogs}. The leading logarithms for the cross sections
are therefore completely determined by a one-loop evaluation, which we
briefly review below. The eikonal cross-section, up to NLO and in momentum
space, can be written as~\footnote{Our presentation is motivated by that of 
Ref.~\cite{Catani:1989ne}.}
\beq
  {\cal S} \!\left( z, Q^2, \epsilon \right) \, = \, 
  \left( 1 + {\cal S}^{(1)}_{\rm virtual} \right) \delta(1 - z)
  + {\cal S}^{(1)}_{\rm real} (z) + {\cal O} \left( \alpha_s^2 \right) \, .
\label{NLOeik}
\eeq
The real radiation contribution can be obtained from the graphs of
Figure~\ref{fig:NLOdiags} using eikonal Feynman rules, and one finds
\beq
  {\cal S}^{(1)}_{\rm real} (z) \, = \, 
  \mu^{2 \epsilon} g_s^2 \, C_F \int \frac{d^d k}{(2 \pi)^{d - 1}} \, 
  \delta_+ (k^2) \, \delta \!\left( 1 - z - \frac{2 k \cdot (p_1 + p_2)}{\hat{s}} 
  \right) \frac{2 p_1 \cdot p_2}{p_1\cdot k \, p_2 \cdot k} \, .
\label{eikxsec}
\eeq
The virtual contribution at ${\cal O}(\alpha_s)$ can be obtained by direct
calculation, or by imposing the soft gluon unitarity requirement
\beq
  \int_0^1 d z \, {\cal S} \!\left( z, Q^2, \epsilon \right) \, = \, 1 \, ,
\label{unitarity}
\eeq
reflecting the requirement that soft divergences from the virtual and real
contributions must cancel, and the fact that Wilson line correlators are pure 
counterterms in dimensional regularisation. This requirement implies
\beq
  {\cal S}^{(1)}_{\rm virtual} \, = \, - \int_0^1 d z \, {\cal S}^{(1)}_{\rm real} (z) \, ,
\label{wvirtual}
\eeq
so that the eikonal cross-section at ${\cal O}(\alpha_s)$ 
can be written as
\beq
  {\cal S}^{(1)}(z) \, = \, \mu^{2\eps} g_s^2 \, C_F 
  \int \frac{d^d k}{(2\pi)^{d - 1}} \, \bigg[ \delta^+(k^2) \,  
  \delta \left(1 - z - \frac{2 k \cdot (p_1 + p_2)}{\hat{s}} \right)
  - \delta(1 - z) \bigg] \frac{2 p_1 \cdot p_2}{p_1\cdot k \, p_2 \cdot k} \, .
\label{eikNLO}
\eeq
To carry out the momentum integral, it is particularly convenient to 
introduce the {\it Sudakov decomposition}
\beq
  k^\mu \, = \, k_{+} \beta_1^\mu + k_{-} \beta_2^\mu + k_{T}^\mu \, , 
\label{Sudakov}
\eeq
where $k_{T}$ is a four-vector transverse to $\beta_1^\mu$ and $\beta_2^\mu$,
\beq
  k_T \cdot \beta_1 \, = \, k_T \cdot \beta_2 \, = \, 0 \, .
\label{kt}
\eeq
Contracting \eq{Sudakov} with $\beta_1^\mu$ and $\beta_2^\mu$, it is
straightforward to verify that the components $k_\pm$ are given by
\beq
  k_+ \, = \, \frac{2 p_2 \cdot k}{\sqrt{\hat{s}}} \, , \qquad 
  k_- \, = \, \frac{2 p_1 \cdot k}{\sqrt{\hat{s}}} \, ;
\label{kpm}
\eeq
furthermore, the integration measure in \eq{eikNLO} becomes
\beq
  \int d^d k \, =  \, \frac{1}{4} \int dk_+ \, dk_- \, d \vec{k}_T^2 \, d \Omega_{d - 2} 
  \left( \vec{k}_T^2 \right)^{(d - 4)/2} \, ,
\label{measure}
\eeq
where $d \Omega_m$ is the element of solid angle in $m$ 
spatial dimensions. \eq{eikNLO} then becomes
\beq
  {\cal S}^{(1)}(z) \, = \, \frac{\mu^{2\eps} \Omega_{d - 2}}{(2 \pi)^{d - 1}} \,
  g_s^2 \, C_F  \int dk_+ \, dk_- \, \left(k_+ k_-\right)^{\frac{d-6}{2}}
  \bigg[ \delta \left(1 - z - \frac{k_+ + k_-}{\sqrt{\hat s}}\right) - \delta(1-z) \bigg] \, .
\label{eikNLO2}
\eeq
The remaining integrals can be easily carried out using
\beq
  k_+ \, = \, \sqrt{\hat{s}} (1 - z) y \, , \qquad 
  k_- \, = \, \sqrt{\hat{s}} (1 - z)(1 - y) \, .
\label{kpmtrans}
\eeq
Taking into account also that 
\beq
  \Omega_{d - 2} \, = \, \frac{2 \pi^{\frac{d - 2}{2}}}{\Gamma(\frac{d - 2}{2})}
\label{Omegad}
\eeq
one has 
\beq
  {\cal S}^{(1)}(z) \, =  \, \frac{\as C_F}{\pi} \left(\frac{\bar \mu^2}{\hat s} \right)^\eps
  \frac{e^{\eps \gamma_E} \Gamma^2 (- \eps)}{\Gamma(1 - \eps)\Gamma(- 2 \eps)} 
  \bigg[ (1 - z)^{- 1 - 2 \eps} + \frac{1}{2 \eps} \delta(1 - z) \bigg] \, ,
\label{eikNLO3}
\ee
where $\bar{\mu}$ is the $\overline{\rm MS}$ renormalisation scale, $\bar{\mu}^2 = 4 
\pi \, {\rm e}^{- \gamma_E} \mu^2$. It is worth noticing at this point that the 
soft function in \eq{eikNLO3} depends on the partonic centre of mass energy 
$\hat s$. Given that in experiments one measures the Drell-Yan cross section 
at fixed $Q$, one  must take into account the fact that $\hat s$ has implicit 
dependence on $z$, and therefore it must be expanded in powers of $z$.
One finds
\beq
  {\cal S}^{(1)}(z) \, = \, \frac{\as C_F}{\pi} \bigg( \frac{\bar \mu^2}{Q^2} \bigg)^{\eps}
  \frac{e^{\eps \gamma_E} \Gamma^2(- \eps)}{\Gamma(1 - \eps)\Gamma(- 2 \eps)} 
  \bigg\{ \Big[ 1 - \eps(1 - z) + \ldots \Big] (1 - z)^{- 1 - 2 \eps} 
  + \frac{1}{2 \eps} \delta (1 - z) \bigg\} \, .
\label{eikNLO4} 
\eeq
It is easy to see that this expansion affects the soft function only beyond leading 
logarithm, and thus we can safely neglect this correction in what follows, and use 
directly \eq{eikNLO3} with the replacement $\hat{s} \to Q^2$. At this 
point, we can safely take the Mellin transform 
\beq
  {\cal S}^{(1)}(N,Q^2) \, = \, \int_0^1 dz \, z^{N-1} \, {\cal S}^{(1)}(z,Q^2) \, ,
\label{MellS}
\eeq
which gives 
\beq
  {\cal S}^{(1)}(N,Q^2) \, = \, \frac{\as C_F}{\pi} 
  \left(\frac{\bar \mu^2}{Q^2}\right)^{\eps} \frac{e^{\eps \gamma_E} 
  \Gamma^2(- \eps)}{\Gamma(1 - \eps)\Gamma(- 2 \eps)} 
  \bigg[ \frac{\Gamma(- 2 \eps)\Gamma(N)}{\Gamma(- 2 \eps + N)} 
  + \frac{1}{2\eps} \bigg] \, .
\label{soft-NLO-Mellin}
\eeq
Expanding in $\epsilon$ one finds
\beqa
  {\cal S}^{(1)} (N, Q^2, \epsilon) & = & 
  \left( \frac{\bar{\mu}^2}{Q^2} \right)^{\epsilon} 
  \frac{\alpha_s}{\pi} \, C_F \left[ \frac{2}{\epsilon} 
  \Big( \psi^{(0)}(N) + \gamma_E \Big) \right. \nonumber \\
  && \left. \quad + \, \frac{6 \psi^{(0)} (N) \left( \psi^{(0)}(N) + 2 \gamma_E \right) 
  - 6 \psi^{(1)}(N) + \pi^2 + 6 \gamma_E^2}{3} \right] \, ,
\label{eikNLO6}
\eeqa  
where $\psi^{(n-1)}$ denotes the $n$-th derivative of the logarithm 
of the $\Gamma$ function. Keeping the dominant behaviour 
as $N \to \infty$ one finds the simple result
\beq
   \left. {\cal S}^{(1)}(N, Q^2, \epsilon) \right|_{\rm LL} \, = \, 
  \left(\frac{\bar{\mu}^2}{Q^2}\right)^{\epsilon} \frac{2\alpha_s}{\pi} \, C_F
  \left[ \frac{\log N}{\eps} + \log^2 N \right] \, , 
\label{eikNLO7}
\eeq
where we kept the leading power of the logarithm separately for the 
divergent and for the finite contributions. As discussed above, we may 
exponentiate this result to obtain the leading logarithmic behaviour at 
all orders. Upon doing so, we may absorb the resulting collinear poles 
into the parton distributions, using the $\overline{\rm MS}$ scheme. 
This amounts to defining renormalised and resummed quark distributions 
via
\beq
  q_{\rm LL} (N, Q^2) \, = \, q (N, Q^2) \exp \left[ \frac{\alpha_s}{\pi} \, 
  C_F \, \frac{\log N}{\epsilon} \right] \, ,
\label{qLL}
\eeq
and similarly for the antiquark, so that \eq{sigmahadN} becomes
\beq
  \left. \int_0^1 d \tau \, \tau^{N - 1} \, \frac{d \sigma_{\rm DY}}{d \tau}
  \right|_{\rm LL} \, = \, \sigma_0 (Q^2) \, q_{\rm LL} (N,Q^2) \,
  \bar{q}_{\rm LL} (N,Q^2) \exp \left[ \frac{2\alpha_s}{\pi}  \, C_F \,
  \log^2 \! N \right] \, .
\label{eikresum}
\eeq
This formula explicitly sums up leading logarithms in $N$ to all orders. 
It can easily be verified that \eq{eikresum} reproduces the well-known 
results of earlier studies, see for example~\cite{Sterman:1987aj,
Catani:1989ne,Moch:2009hr}, both in Mellin space and in momentum 
space. We note in passing that in our analysis that the dimensional 
regularisation scale $\mu$ appears only through the factor $\mu^{2 \epsilon}$, 
as must be the case on dimensional grounds. Given that $\mu$ is then
identified with the renormalisation and factorisation scales, it follows
that logarithms of these scales (which may be chosen to depend on $z$)
must be suppressed by a single power of $\epsilon$, and thus do not
contribute to the leading logarithmic behaviour in the threshold
variable $(1 - z)$, as could be expected. The same argument will hold 
at NLP level. We also note that, in going beyond LP level, we will have 
to keep track of subleading terms in \eq{eikNLO6}. Expanding this to NLP 
order one finds 
\beq 
  {\cal S}^{(1)} (N, Q^2, \epsilon) \, = \, \left( \frac{\bar{\mu}^2}{Q^2}
  \right)^{\epsilon} \frac{2 \alpha_s C_F }{\pi} \left[ \frac{1}{\epsilon}
  \left(\log N - \frac{1}{2N} \right) + \log^2 N - \frac{\log N}{N} \right] \, ,
\label{eikNLO8}
\eeq
which will be useful later on.


\section{Threshold resummation at next-to-leading power}
\label{sec:NLP}

In the previous section, we reviewed the exponentiation of leading logarithmic 
threshold contributions to the Drell-Yan cross-section at leading power. We 
now discuss how to extend this procedure to next-to-leading power, and we 
will keep our remarks general enough to apply to both quark and gluon-initiated 
processes, and for general colour-singlet final states. Recall that LP resummation
at LL accuracy relied on two facts: the exponentiation of the soft function before 
integration over phase space (at squared matrix element level), and the
factorisation of phase space for $m$ parton emissions into $m$ decoupled 
single-parton phase space integrals. This motivates the following schematic 
decomposition of the partonic cross-section up to NLP order, which was already 
shown to be useful in Ref.~\cite{Laenen:2010uz}:
\beq
   \hat \sigma \, = \, \frac{1}{2 \hat{s}} \left[ \, \int d \Phi_{\rm LP} 
  \left| {\cal M} \right|_{\rm LP}^2 + \int d \Phi_{\rm LP} 
  \left| {\cal M} \right|_{\rm NLP}^2 + \int d \Phi_{\rm NLP}
  \left| {\cal M} \right|_{\rm LP}^2 + \, \ldots  \right] \, .
\label{sigmaNLP}
\eeq
The first term on the right-hand side of \eq{sigmaNLP} gives the leading-power 
result of \secn{sec:LP}, integrating the leading-power squared matrix element 
with leading-power phase space, {\it i.e.} neglecting correlations between 
radiated partons. The second term consists of the NLP contribution to the 
squared matrix element, integrated with LP phase space. The third term 
consists of the LP matrix element, but where the phase space includes the 
effect of parton correlations at NLP. Finally, the ellipsis denotes terms which
are NNLP and beyond in the threshold expansion. Based on this classification, 
the task of determining whether LL NLP terms can be resummed amounts 
to elucidating the relevant structure of the NLP matrix element, as well as 
considering whether NLP corrections to the LP phase space are significant. 
Let us consider each of these issues in turn.


\subsection{Structure of the NLP squared matrix element at LL}
\label{sec:NLPmatrix}

\begin{figure}
\begin{center}
\scalebox{0.8}{\includegraphics{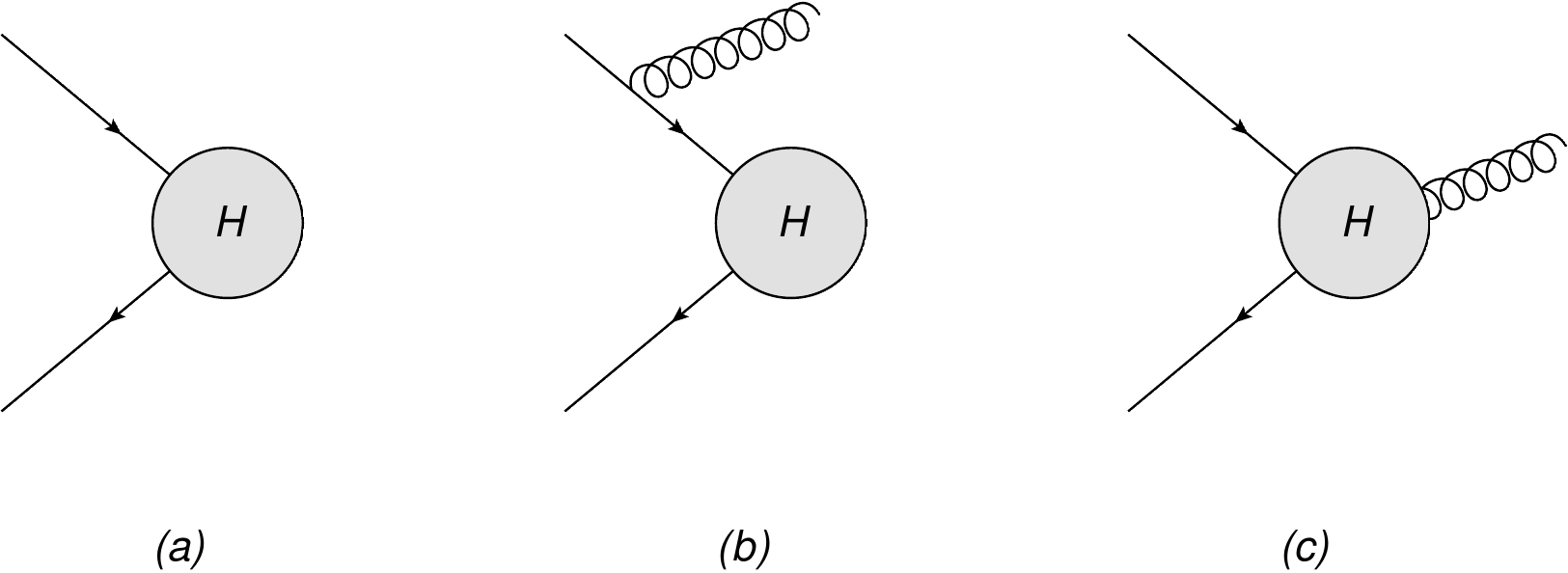}}
\caption{(a) Non-radiative amplitude with two incoming particles and a
  {\it hard interaction} ${\cal H}$; (b) external emission
  contribution; (c) internal emission contribution.}
\label{fig:loopamp}
\end{center}
\end{figure}

For simplicity, let us first describe the structure of squared matrix 
elements at NLP level when the hard emitters are massive, following 
Refs.~\cite{Laenen:2008gt,Laenen:2010uz} (themselves building 
on Refs.~\cite{Low:1958sn,Burnett:1967km}). In Figure~\ref{fig:loopamp}(a), 
we draw a non-radiative amplitude with an incoming quark and antiquark, 
which interact via a {\it hard interaction} ${\cal H}$. Radiation can then 
be divided into two types of contribution:
\begin{enumerate}
\item {\it External emissions}. In this case, radiation couples directly to 
  the incoming hard lines, as exemplified in figure~\ref{fig:loopamp}(b). 
  Notice that this case includes all radiation that does not resolve the structure
  of the hard interaction, and incorporates intricate diagrammatic cancellations
  that lead to a factorised form of the amplitude: indeed, all radiation at
  leading power falls in this category.
\item {\it Internal emissions}. At next-to-leading power, non-factorisable
  contributions from next-to-soft partons arise, which can be depicted as
  originating from inside the hard interaction, as in figure~\ref{fig:loopamp}(c). 
  This corresponds to the insertion of sub-leading power operators in an
  effective field theory language, and it is the first level of interaction where
  soft radiation begins to unravel the structure of the hard scattering.
\end{enumerate}
As shown for the first time in Ref.~\cite{Laenen:2008gt}, external emissions 
can be described by {\it generalised Wilson lines}, which extend the definition
given in \eq{Phidef} to next-to-leading power in the soft expansion. Along the 
lines of \eq{Wilsonexp}, we may write this operator in momentum space 
as~\cite{Bonocore:2016awd}
\beq
  F (p) \, = \, {\cal P} \exp \left[ g_s {\bf T}^a \int \frac{d^d k}{(2 \pi)^d} \, 
  \tilde{A}_\mu^a(k) \left(\frac{p^\mu}{p \cdot k} - \frac{k^\mu}{2 p \cdot k}
  + k^2 \frac{p^\mu}{2 (p \cdot k)^2} + 
  {\rm i} k_\nu \frac{S^{\nu \mu}}{p \cdot k}\right) + \ldots \right] \, 
\label{fidef}
\eeq
for a generalised semi-infinite straight Wilson in the direction of 
four-momentum $p$. Here ${\bf T}^a$ is a colour generator in the
appropriate representation, and $S^{\mu\nu}$ is the generator of Lorentz
transformations for the parton under consideration, vanishing for scalar 
fields, while it is given by
\beq
  \left( S^{\nu\mu} \right)_{\alpha \beta} \, \equiv \, {\Sigma^{\mu \nu}}_{\alpha \beta} 
  \, = \, \frac{{\rm i}}{4} \left[ \gamma^\nu, \gamma^\mu \right]_{\alpha \beta}
\label{Sigmadef}
\eeq
for spin-$1/2$ fields, and 
\beq
  \left( {S^{\nu\mu}} \right)_{\rho \sigma} \, \equiv \, {M^{\nu \mu}}_{\rho \sigma}
  \, = \,  {\rm i} \left( \delta^\nu_\rho \, \delta^\mu_\sigma - \delta^\nu_\sigma \, 
  \delta^\mu_\rho \right)
\label{Mdef}
\eeq
for vector fields. The first term in \eq{fidef} corresponds to the eikonal Feynman 
rule of \eq{Wilsonexp}, and the remaining terms (suppressed by one power of 
the gluon momentum $k$) correspond to effective {\it next-to-eikonal} Feynman 
rules, describing the emission of {\it next-to-soft} gauge bosons~\cite{Laenen:2008gt}.
The ellipsis in \eq{fidef} refers to terms involving two gluons being emitted from 
the same point along the Wilson line, through seagull-type vertices. These vertices 
start contributing to the cross section at NNLO (either through 
double radiation, or one-loop corrections to single radiation), therefore, 
under mild assumptions, they cannot contribute at leading logarithmic accuracy
at NLP (as was the case at LP). Indeed, our proposed resummation rests upon 
an amplitude-level factorisation theorem~\cite{DelDuca:1990gz,Bonocore:2015esa,
Bonocore:2016awd}, which implies the existence of evolution equations, which 
in turn can be understood in terms of renormalisation of suitable operator matrix 
elements: the solution of evolution equations of this type always leads to a 
non-trivial pattern of exponentiation, with single logarithms in the exponent 
generating double logarithms in the cross section. Such a pattern of exponentiation
implies that all leading logarithms are generated by the one-loop exponent. A test 
of this argument is provided by Ref.~\cite{Laenen:2008ux}, where the exponentiation
of leading logarithms at NLP was explictly tested at NNLO; finally, as 
a further check, we verify in Appendix C that next-to-soft Feynman rules for 
double radiation in \eq{fidef} do no contribute to leading logarithms in the case 
of Drell-Yan production at two loops.

Let us now consider the contribution of internal emissions. When massive 
external particles are being considered, the hard interaction is analytic in
the total momentum $K$ of the emitted radiation, and can safely be expanded
about the soft limit $K^\mu \rightarrow 0$. One may then show, using Ward 
identities, that the effect of a single internal emission is given by derivatives 
of the non-radiative amplitude with respect to its external momenta. As has 
been noted in the context of the so-called {\it next-to-soft theorems} of
Refs.~\cite{Casali:2014xpa,Cachazo:2014fwa}~\footnote{See
  Ref.~\cite{White:2014qia} for a discussion of how to relate the more
  formal works of Refs.~\cite{Casali:2014xpa,Cachazo:2014fwa} to the
  present framework.}, these derivatives can be organised in terms of the 
orbital angular momentum operator associated with each external leg, 
which, in momentum space, has the form
\beq
  L^{(i)}_{\nu \mu} \, = \, {\rm i} \! \left(p_{i \nu} \frac{\partial}{\partial p_{i \mu}}
  - p_{i \mu} \frac{\partial}{\partial p_{i \nu}} \right) \, .
\label{Lmunudef}
\eeq
On each hard leg, this combines with the spin angular momentum contribution
to construct the total angular momentum operator
\beq
  S^{\nu \mu} \, \rightarrow \, S^{\nu \mu} + L^{\nu \mu} \, \equiv \, 
  J^{\nu \mu} \, .
\label{jmunu}
\eeq
In Ref.~\cite{Laenen:2008gt}, the orbital angular momentum contribution 
was not included in the generalised Wilson line operator of \eq{fidef}, despite 
the fact that it might make sense to do so, given that the internal and external 
emission contributions are not separately gauge-invariant, but instead combine 
into a gauge-invariant object, the total angular momentum. For practical
purposes, however, it remains convenient to keep the orbital angular momentum 
separate, given that it involves derivatives which have yet to act on the hard 
interaction. How to keep track of such contributions will be discussed explicitly 
below.

\begin{figure}
\begin{center}
  \scalebox{0.6}{\includegraphics{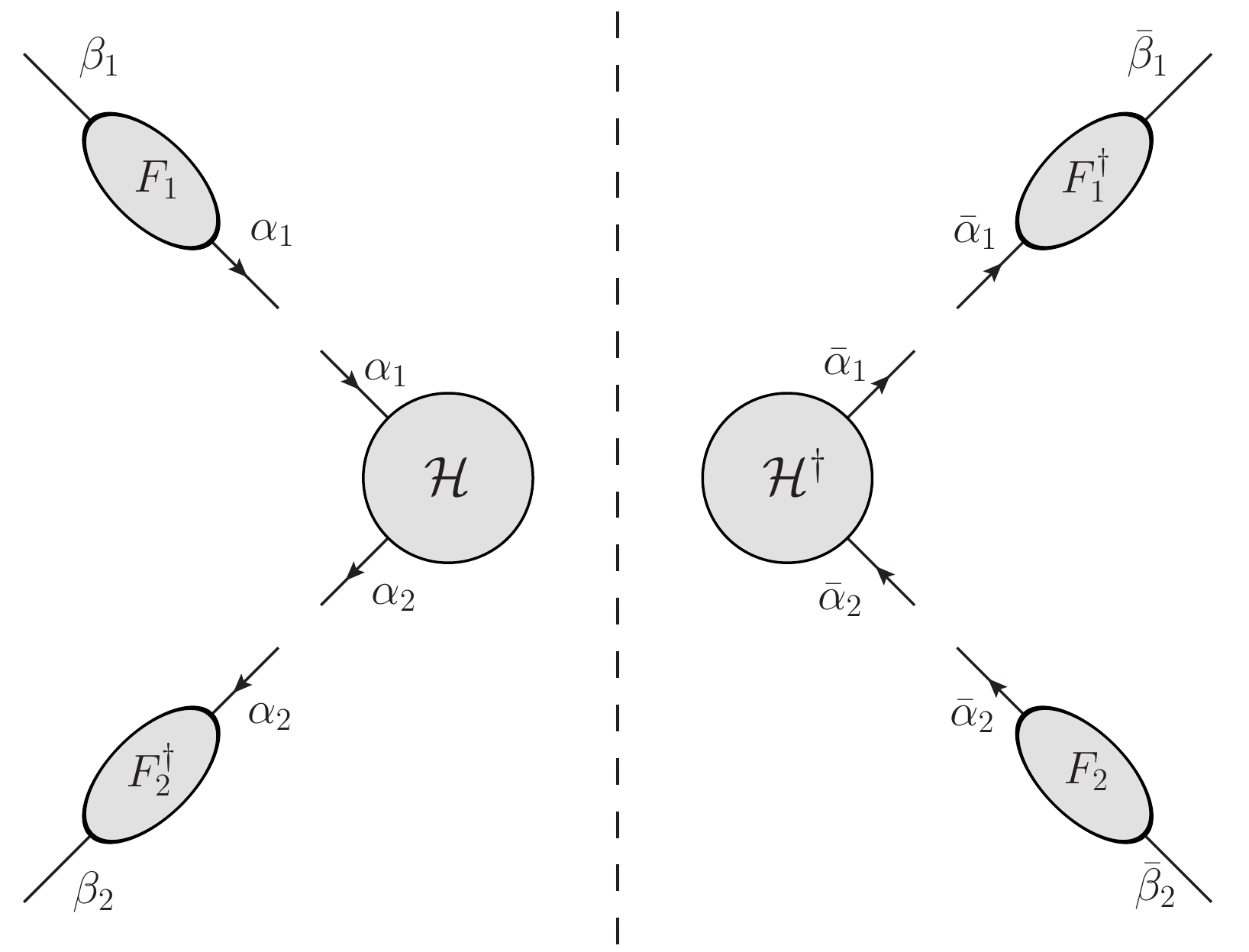}}
  \caption{Labelling of spin indices for the squared amplitude, where
  $H$ is the hard function, and $F_i$ a generalised Wilson line.}
\label{fig:spinindices}
\end{center}
\end{figure}

Armed with the operator defined in \eq{fidef}, we may construct a {\it next-to-soft} 
function by analogy with the LP soft function of \eq{Sdef}, as
\beq
  \widetilde{\cal S} \left( z, Q^2, \epsilon \right) \, = \, \frac{1}{N_c} 
  \sum_{n, \, {\rm LP}} {\rm Tr} \left[ \big \langle 0 \big|
  F^\dag (p_1) F (p_2) \big| n \big\rangle
  \big\langle n \big| F^\dag (p_2) F (p_1) \big| 0 \big\rangle \right]
  \delta \bigg( z - \frac{Q^2}{\hat{s}} \bigg) \, .
\label{Stildedef}
\eeq
Here we have replaced the Wilson line operators in the LP soft function 
by their NLP counterparts, and the subscript in the sum over final 
states indicates that all phase space integrals are to be carried out 
with LP phase space only ({\it i.e.} with a measure of integration consisting
of a product of single-gluon phase space integrals). Corrections to this will 
be considered in \secn{sec:NLPphasespace}. Note also that all generalised 
Wilson lines are semi-infinite straight lines proceeding from the origin in 
position space. At NLP accuracy the cross section is sensitive to a potential 
non-zero initial position, but this is related to the derivative contributions 
above~\cite{Laenen:2008gt}, which are to be dealt with separately. 
As was the case at LP, the next-to-soft function in \eq{Stildedef} can 
be shown to exponentiate using replica trick arguments (see 
Appendix~\ref{app:replica}). At NLP, however, we must carefully
disentangle what this means, given that the generalised Wilson line 
of \eq{fidef} is matrix-valued in the spin space of the external hard
particles. As an example, consider the spin-$1/2$ case, and let us 
write the non-radiative amplitude with an incoming fermion and antifermion 
with explicit spin indices $\{ \alpha, \beta \}$, as
\beq
  {\cal M} \, = \, \bar{v}_\alpha (p_2) M^{\alpha \beta} u_\beta(p_1) \, ,
\label{A0def}
\eeq
so that the spin matrix $M$ is defined by stripping off the initial state wave 
functions from the full amplitude. The next-to-soft function of \eq{Stildedef} 
can then be explicitly written as a spin operator
\beq
  \widetilde{\cal S}^{\,\, \alpha_1 \alpha_2 \bar{\alpha}_1 \bar{\alpha}_2}_{\beta_1 
  \beta_2 \bar{\beta}_1 \bar{\beta}_2} (z, Q^2, \epsilon) \, = \, 
  \frac{1}{N_c} \sum_{n, \, {\rm LP}}
  \big\langle 0 \big| F^{\dag \, \bar{\alpha}_1}_{\bar{\beta}_1} (p_1) \,
  F^{\, \bar{\alpha}_2}_{\bar{\beta}_2} (p_2) \big| n \big\rangle 
  \big\langle n \big| F^{ \dag \, {\alpha_2}}_{\beta_2} (p_2) \,
  F^{\, \alpha_1}_{\beta_1} (p_1) \big| 0 \big\rangle \, 
  \delta \bigg( z - \frac{Q^2}{\hat{s}} \bigg) \, ,
\label{Stildedef2}
\eeq
where the ordering of spinor indices is depicted in Figure~\ref{fig:spinindices}.
The spin matrix $M^{\alpha \beta}$ in \eq{A0def} factorises into a product of 
hard and next-to-soft factors, so that the integrated squared matrix element, 
dressed by arbitrary amounts of radiation from the next-to-soft function, can be
written as
\beq
  \int d \Phi^{(n + 1)} \left| {\cal M} \right|^2 \, = \, 
  \widetilde{\cal S}^{\,\, \alpha_1 \alpha_2 \bar{\alpha}_1 \bar{\alpha}_2}_{\beta_1 
  \beta_2 \bar{\beta}_1 \bar{\beta}_2} (z, Q^2, \epsilon)
  \int d \Phi^{(1)}
  \Big[\bar{v}^{\beta_2} (p_2) {\cal H}_{\alpha_1 \alpha_2} u^{\beta_1} (p_1) \Big]
  \Big[\bar{u}^{\bar{\beta}_1} (p_1) {\cal H}^{\dag}_{\bar{\alpha}_1 
  \bar{\alpha}_2} v^{\bar{\beta}_2} (p_2) \Big] \, ,
\label{SH}
\eeq 
where $d \Phi^{(m)}$ denotes the $m$-particle Lorentz-invariant phase space 
measure, and the integration over the phase space of the heavy vector boson 
has been singled out, relying upon the  factorisation of the $n$-body phase 
space at LP. An expression very similar to \eq{SH} holds for incoming particles
of spin one, with spinors replaced by polarisation vectors, and spinor indices 
by vector indices.

The discussion so far applies strictly only to the case of massive
external particles. When massless particles are involved, it is no
longer true that the hard function ${\cal H}$ is analytic in the
momentum carried by soft radiation: it develops logarithmic
singularities due to the presence of collinear divergences. As
discussed in the Introduction, this was first explored in a QED
context in Ref.~\cite{DelDuca:1990gz}, which presented a factorisation
formula at amplitude level, extending the Low-Burnett-Kroll theorem to
include collinear effects. Similar ideas have recently been extended
to full QCD~\cite{Bonocore:2015esa,Bonocore:2016awd}, and analysed
using
SCET~\cite{Larkoski:2014bxa,Beneke:2017ztn,Beneke:2018rbh,Beneke:2018gvs,Kolodrubetz:2016uim,Moult:2016fqy,Moult:2017rpl,
  Feige:2017zci,Chang:2017atu,Boughezal:2016zws,Boughezal:2018mvf,Moult:2017jsg,Ebert:2018lzn,Ebert:2018gsn,Moult:2018jjd,Bhattacharya:2018vph},
while an alternative, first-principles, diagrammatic approach has been
explored in Ref.~\cite{Gervais:2017yxv}. Common to all these
approaches is the factorisation of collinear contributions into
universal functions, sensitive to the spin of the colliding particles
but otherwise independent of the details of the hard scattering. More
precisely, one may recall that, at leading power, collinear radiation
is accounted for by means of jet-type functions, such as the parton
distributions introduced in \eq{Deltafac}. At NLP, one must generalise
this analysis, expressing radiative amplitudes in terms of new types
of jet functions, describing soft emissions from collinearly enhanced
configurations. The first such {\it radiative jet function} was
proposed in Ref.~\cite{DelDuca:1990gz}, and was recently calculated at
one loop in QCD for external quarks in Refs~\cite{Bonocore:2015esa,
  Bonocore:2016awd}, where it was used to reproduce known NLP
threshold logarithms in Drell-Yan production at NNLO. The
amplitude-level factorisation proposed in~\cite{Bonocore:2015esa} is
expected to apply only to annihilation processes involving two
colliding hard partons producing colourless final states: on the other
hand, the analysis of Ref.~\cite{Gervais:2017yxv} (which focuses on
scalar theories, but considers more general scattering processes), and
the results of
Refs.~\cite{Larkoski:2014bxa,Beneke:2017ztn,Beneke:2018rbh,Beneke:2018gvs} 
suggest that further types of jet emission function are necessary in QCD,
which have yet to be calculated.

Importantly, for the purposes of the present paper, radiative jet functions
can be ignored, since it can be shown that leading logarithms at NLP can
only arise from momentum regions of integration that are already fully 
accounted for by the next-to-soft function introduced in \eq{fidef}. To illustrate
this point, consider first, for comparison, the well-understood situation at leading 
power\footnote{For clarity, we focus here on real-radiation contributions to the 
inclusive cross section, which are the origin of the $z$-dependence we are interested 
in.}. In that case, threshold singularities, inducing non-analytic behaviour 
at $z \to 1$, are directly related to infrared singularities of the amplitude; 
these, in turn, arise from integrations of the relevant momentum components 
(`normal variables') near singular surfaces in momentum space, which can 
be completely characterised to all orders in perturbation theory by means of 
Landau equations and power counting techniques~\cite{Sterman:1995fz}. 
For massless theories, it can be shown in general that infrared singularities 
arise only from soft and collinear momentum configurations. At leading power, 
therefore, one finds that at $n$ loops there are precisely $2 n$ normal variables
that must be integrated with a logarithmic measure: in a suitable frame, these
can be taken to be $n$ parton energies $E_i$, with a leading-power integration
measure $d E_i/E_i$, and $n$ transverse momenta with respect to the directions 
defined by external particles, $k_{i T}$, with a leading-power integration measure 
$d k_{i T}/k_{i T}$. Threshold logarithms in general arise when different 
combinations of normal variables become small at different rates, but leading 
logarithms arise only with a very specific scaling, when all energies and transverse 
momenta are strongly ordered, say $E_1 \ll \ldots \ll E_n$ and $ k_{1 T} \ll \ldots \ll  
k_{n T}$. In that limit, at LP, the $2 n$ logarithmic integrations yield contributions 
of the form $\ln^{2 n - 1} (1 - z)/(1-z)$, since the last logarithmic integration must not 
performed when computing $d \hat{\sigma}/d z$. At NLP, either the phase-space 
measure or the squared matrix element provide a single power of one of the normal
variables, so that only $2 n - 1$ momentum components need be integrated with 
a logarithmic measure. Once again, leading logarithms will arise from the configuration 
where the remaining normal variables are strongly ordered, with $2 n - 1$ integrations 
leading to contributions of the form $\ln^{2 n - 1} (1 - z)$, while the $z$ integration 
will not introduce further singularities at NLP. Now, two possibilities arise. On the one 
hand, the normal variable whose integration has become non-singular can be 
a transverse momentum, in which case the corresponding parton is soft, but 
not strictly collinear: this configuration is accounted for by the leading-power
soft function, which contains `wide-angle' soft gluons. On the other hand, the 
suppressed variable can be an energy: in this case, all transverse momenta
must be strongly ordered; such next-to-soft, collinear configurations are 
accounted for by the next-to-soft function defined in \eq{Stildedef}. Notice
that radiative jet functions such as the one computed in~\cite{Bonocore:2015esa}
also contain the next-to-soft, collinear configuration: this, however, contributes to 
a double counting that must be explicitly subtracted, either by introducing eikonal
jets, as done in \eq{Deltafac}, or by defining an appropriate counterterm,
as done for example in Ref.~\cite{Bonocore:2016awd}. The subtracted
radiative jet function then contains only hard collinear configurations for
all radiated partons, and cannot contribute at leading logarithmic accuracy.

An explicit example and test of the above discussion is provided in
Ref.~\cite{Bonocore:2016awd}, where the non-abelian radiative jet
function for quarks was computed at one-loop order, and the overlap
between (next-to-)soft and collinear emissions was explicitly
identified. Furthermore, a large class of (N)LP threshold effects has
been calculated in Drell-Yan production at
NNLO~\cite{Bonocore:2014wua} and N$^3$LO~\cite{Bahjat-Abbas:2018hpv}
using the {\it method of
  regions}~\cite{Pak:2010pt,Jantzen:2011nz,Jantzen:2012mw}, which
allows for a precise identification of the (next-to)soft and/or
collinear origin of all contributions to the cross section. The role
of hard collinear effects is indeed found in these studies to be
associated with NLL terms and beyond~\footnote{The fact that hard
  collinear contributions are subleading has also been argued in
  various SCET
  approaches~\cite{Beneke:2018gvs,Moult:2016fqy,Moult:2017jsg,Bhattacharya:2018vph}.},
while all LL contributions can be traced to the (next-to)soft
function, if the results are recast in the present framework. Notice
that, as discussed above, upon exponentiation leading logarithms at
NLP must be generated by one-loop contributions: the results
of~\cite{Bonocore:2016awd,Bonocore:2014wua, Bahjat-Abbas:2018hpv}
therefore provide a complete test of our argument for the Drell-Yan
process.

To summarise, NLP contributions to squared matrix elements can be categorised 
into two main types, as follows.
\begin{enumerate}
\item[(i)] {\it (Next-to-)soft emissions}. These are captured by the next-to-soft 
  function, defined in terms of generalised Wilson lines in \eq{Stildedef}, together 
  with the orbital angular momentum contributions associated with internal emissions 
  in Figure~\ref{fig:loopamp}.  As at LP, the next-to-soft function exponentiates 
  (see Appendix~\ref{app:replica}).
\item[(ii)] {\it Collinear contributions}. These are described by radiative jet functions, 
  which overlap with the next-to-soft function. Upon removing the double counting, 
  the remaining collinear effects do not contribute at LL accuracy.
\end{enumerate}
In this section, we have discussed the second term on the right-hand side of 
\eq{sigmaNLP}, and argued that the next-to-soft function underlies all contributions 
to the NLP matrix element that can result in LL terms in the cross-section. This is 
only part of the story: we must also check whether or not LL terms can arise from 
the LP matrix element, once correlations between radiated gluons (a NLP
effect) are included. This is the subject of the following section.


\subsection{Corrections to the LP phase space}
\label{sec:NLPphasespace}

The third term in \eq{sigmaNLP} consists of the LP matrix element integrated 
over the NLP phase space. To see whether or not this term can give LL
contributions at NLP, it is sufficient to take the LL contribution to the LP matrix 
element at each order, and then to evaluate the phase space integral up 
to NLP order. The LL contributions to the matrix element have already been 
discussed in \secn{sec:LP}, and involve exponentiating the NLO eikonal 
squared matrix element. This generates terms with $n \geq 1$ gluon
emissions, and, according to \eq{Sdef}, one must then integrate each 
such term over the $n$-gluon phase space. Considering all possible 
contributions to an $n$-gluon final state yields a squared matrix element
of the form
\beq
  | {\cal M} |^2_{\rm LP, n} \, = \, f \big( \alpha_s, \epsilon, \mu^2 \big) \,
  \prod_{i = 1}^n \frac{p_1\cdot p_2}{p_1\cdot k_i \, p_2 \cdot k_i} \, ,
\label{Mndef}
\eeq
where the prefactor $f(\alpha_s,\epsilon,\mu^2)$ collects coupling 
dependence, possible poles in $\epsilon$ due to the integration over loop
momenta, and combinatorial factors from the exponentiation of the
squared matrix element. The explicit form of this function is irrelevant 
for what follows. We must now integrate \eq{Mndef} over the $(n+1)$-body 
phase space, consisting of $n$ gluons, as well as the electroweak vector 
boson that defines the final state at LO. The integration measure is given by
\beq
  d \Phi^{(n+1)} \, = \, \bigg[ \prod_{i = 1}^n \int \frac{d^d k_i}{(2 \pi)^{d - 1}}
  \, \delta_+ (k_i^2) \bigg] \delta \bigg(1 - z - 2 \sum_{i=1}^n 
  \frac{k_i \cdot (p_1 + p_2)}{\hat{s}} + 2 \sum_{i  <  j} \frac{k_i \cdot k_j}
  {\hat{s}} \bigg) \, ,
\label{dPhin}
\eeq
where the integration of the vector boson momentum has already been 
carried out, using the overall momentum conservation $\delta$ function. 
In order to compute the integral, it is particularly convenient to use the 
Sudakov decomposition of \eq{Sudakov} for each momentum $k_i$.
One finds
\beqa
  \int d \Phi^{(n+1)} | {\cal M} |^2_{\rm LP, n} & = & f (\alpha_s,\epsilon,\mu^2)
  \left[ \prod_{i = 1}^n \frac{1}{\pi \hat{s}}
  \int_0^\infty \frac{d k_{i+}}{k_{i+}} \int_0^\infty \frac{dk_{i-}}{k_{i-}}
  \int \frac{d^{d - 2} \vec{k}_{iT}}{(2 \pi)^{d - 2}} \, \delta( k_{i-} k_{i+} - 
  \vec{k}_{iT}^2) \right] \nonumber \\
  && \times \, 
  \delta \bigg( 1 - z - \frac{(k_{i+} + k_{i-})}{\sqrt{\hat{s}}}
  + 2 \sum_{i< j} \frac{k_i \cdot k_j}{\sqrt{\hat{s}}} \bigg) \, ,
\label{PScalc1}
\eeqa
where we absorbed the $\theta$ functions from the factors $\delta_+(k_i^2)$ 
into the integration limits for $k_{i \pm}$. In order to proceed, we can represent 
the $\delta$ function in the second line of \eq{PScalc1} using
\beq
  \delta (x) \, = \, \int_{- {\rm i} \infty}^{{\rm i} \infty} \, \frac{d T}{2 \pi {\rm i}} \,  
  {\rm e}^{T x} \, .
\label{deltaid}
\eeq
We may then rewrite \eq{PScalc1} as 
\beqa
  \int d \Phi^{(n+1)} | {\cal M} |^2_{\rm LP, n} & = & f (\alpha_s, \epsilon, \mu^2)
  \int_{-{\rm i} \infty}^{{\rm i} \infty} \frac{d T}{2 \pi {\rm i}} \, {\rm e}^{T (1 - z)}
  \Bigg[ \prod_{i = 1}^n \frac{1}{\pi \hat{s}}
  \int_0^\infty \frac{dk_{i+}}{k_{i+}}\int_0^\infty \frac{dk_{i-}}{k_{i-}} 
  \label{PScalc2} \\
  & \times & \!\!\!
  \int \frac{d^{d - 2} \vec{k}_{iT}}{(2 \pi)^{d - 2}} \, 
  \delta (k_{i-} k_{i+} - \vec{k}_{iT}^2) \, 
  {\rm e}^{- \frac{T (k_{i+} + k_{i-})}{\sqrt{\hat{s}}}} \Bigg]
  \bigg[ 1 + \frac{2 T}{\hat{s}} \sum_{i < j} k_i\cdot k_j
  + {\cal O} (T^2) \bigg] \, , \nonumber
\eeqa
where in the second line we Taylor-expanded the term in the exponent
that is quadratic in soft momentum, anticipating that higher order contributions 
in $T$ in the last line will correspond to subleading powers of $(1 - z)$ in the 
final result. We will verify this fact later, but, for the moment, note that the 
term at ${\cal O}(T)$ corresponds to a phase space correlation between 
pairs of gluons that is absent at LP. Thus, this term constitutes the ``NLP 
phase space'' correction referred to in \eq{sigmaNLP}. In the Sudakov 
decomposition, the dot product of gluon momenta reads
\beq
  k_i \cdot k_j \, = \, \frac{k_{i+} k_{j-} + k_{i-} k_{j+}}{2} - \vec{k}_{iT}
  \cdot \vec{k}_{jT} \, .
\label{kikjsud}
\eeq
The term involving the transverse momenta leads to an odd integrand in
each $\vec{k}_{iT}$ in \eq{PScalc2}, and will therefore give a vanishing contribution
to the final result. We can then carry out the remaining transverse momentum 
integrals to obtain
\beqa
  \int d \Phi^{(n + 1)} | {\cal M} |^2_{\rm LP, n}
  & = & f (\alpha_s,\epsilon,\mu^2) \int_{- {\rm i} \infty}^{ {\rm i} \infty}
  \frac{dT}{2 \pi {\rm i}} \, {\rm e}^{T(1 - z)}
  \Bigg[ \prod_{i = 1}^n \frac{\Omega_{d - 2}}{\hat{s}(2 \pi)^{d - 1}}
  \int_0^\infty dk_{i+} \int_0^\infty dk_{i-}  
  \label{PScalc3} \\
  & & \left( k_{i+} \, k_{i-} \right)^{(d - 6)/2}
  {\rm e}^{- \frac{T (k_{i+} + k_{i-})}{\sqrt{\hat{s}}}} \Bigg]
  \bigg[ 1 + \frac{T}{\hat{s}} \sum_{i < j} \left( k_{i+} k_{j-} + k_{i-} k_{j+} \right)
  + {\cal O}(T^2) \bigg] \, . \nonumber
\eeqa
Next, the $k_{i \pm}$ integrals can be straightforwardly carried out to give 
\beqa
  \int d \Phi^{(n+1)} | {\cal M} |^2_{\rm LP, n}
  & =& f (\alpha_s,\epsilon,\mu^2) \, 
  \frac{\Omega_{d - 2}^n \, \hat{s}^{n (d - 6)/2}}{(2\pi)^{n (d - 1)}}
  \int_{- {\rm i} \infty}^{{\rm i} \infty} \frac{dT}{2 \pi {\rm i}} \, {\rm e}^{T(1 - z)} 
  \label{PScalc4} \\
  & & \hspace{-3.2cm} \times \, \Bigg[ \frac{1}{T^{n (d - 4)}} \, 
  \Gamma^{2 n} \bigg(\frac{d - 4}{2} \bigg)
  + \, \frac{n (n - 1)}{T^{n (d - 4) +1}} \, \Gamma^{2 n - 2}
  \bigg( \frac{d - 4}{2} \bigg) \, \Gamma^2 \bigg( \frac{d - 2}{2} \bigg)
  + {\cal O} \bigg( \frac{1}{T^{n (d - 4) + 2}} \bigg) \Bigg] \, . \nonumber
\eeqa
The integral in $T$ is recognisable as an inverse Laplace transform, which 
yields the result
\beqa
  \int d \Phi^{(n+1)} | {\cal M} |^2_{\rm LP, n}
  & = & f (\alpha_s,\epsilon,\mu^2) \, \frac{\Omega_{d - 2}^n \, 
  \hat{s}^{n (d - 6)/2}}{(2 \pi)^{n (d - 1)}} \, 
  \frac{\Gamma^{2n}[(d-4)/2]}{\Gamma[n(d-4)]} \, (1 - z)^{n (d - 4) - 1}
  \nonumber \\
  & & \hspace{5mm} \times \,
  \bigg\{ 1 + \frac{(n - 1)(d - 4)(1 - z)}{4} + {\cal O} \big[ (1 - z)^2 \big] \bigg\} \, .
\label{PScalcres}
\eeqa
Note that the terms ${\cal O}(T^2)$ we have neglected in expanding the
exponential factor in \eq{PScalc2} give subleading power corrections in $(1-z)$, 
justifying the approximation made above. \eq{PScalcres} is the final result of 
integrating the LP contribution to the matrix element responsible for LL terms,
with the multigluon phase-space measure expanded to NLP order. The second 
term in the last line of \eq{PScalcres} is the desired NLP correction, as can be 
seen by the fact that it is suppressed by a single power of $(1 - z)$. Furthermore,
it contains an explicit factor of $d - 4 = - 2 \epsilon$, which directly implies that 
the phase space correction does not affect LL terms, which are associated with 
the most singular poles in $\epsilon$. 

In summary, we have shown that the third term in \eq{sigmaNLP}, consisting 
of the LP matrix element dressed with NLP phase space corrections, does not 
contribute to LL terms at NLP order. It can thus be neglected for the purposes 
of this paper. Combining this observation with the results of the previous section, 
we now have everything we need to perform an explicit resummation of LL NLP
threshold logarithms in Drell-Yan production. We turn to this task in the next 
section.


\subsection{Resummation of leading NLP logs in Drell-Yan production}
\label{sec:resummation}

In the previous sections, we have seen that LL contributions at NLP
level are governed by next-to-soft radiation. This in turn is captured
by the {\it next-to-soft function} defined in \eq{Stildedef}, possibly
complemented by contributions involving the orbital angular momentum
of each incoming parton. In this section, we apply these ideas to
resum LL NLP terms in Drell-Yan production. While clearly very
interesting for its own sake, this example is also a useful warm-up
case: first, it will allow us to make contact with the LP treatment of
\secn{sec:LP}; furthermore, in this case the hard interaction is
especially simple, so that its derivatives with respect to the
external momenta vanish at leading order. Thus, we do not have to
worry about orbital angular momentum contributions at LL accuracy, and
it is sufficient to calculate the next-to-soft function. Once this has
been calculated for single radiation, it may be exponentiated (as at
LP), yielding the resummation formula that we are seeking.

\begin{figure}
\begin{center}
\scalebox{0.8}{\includegraphics{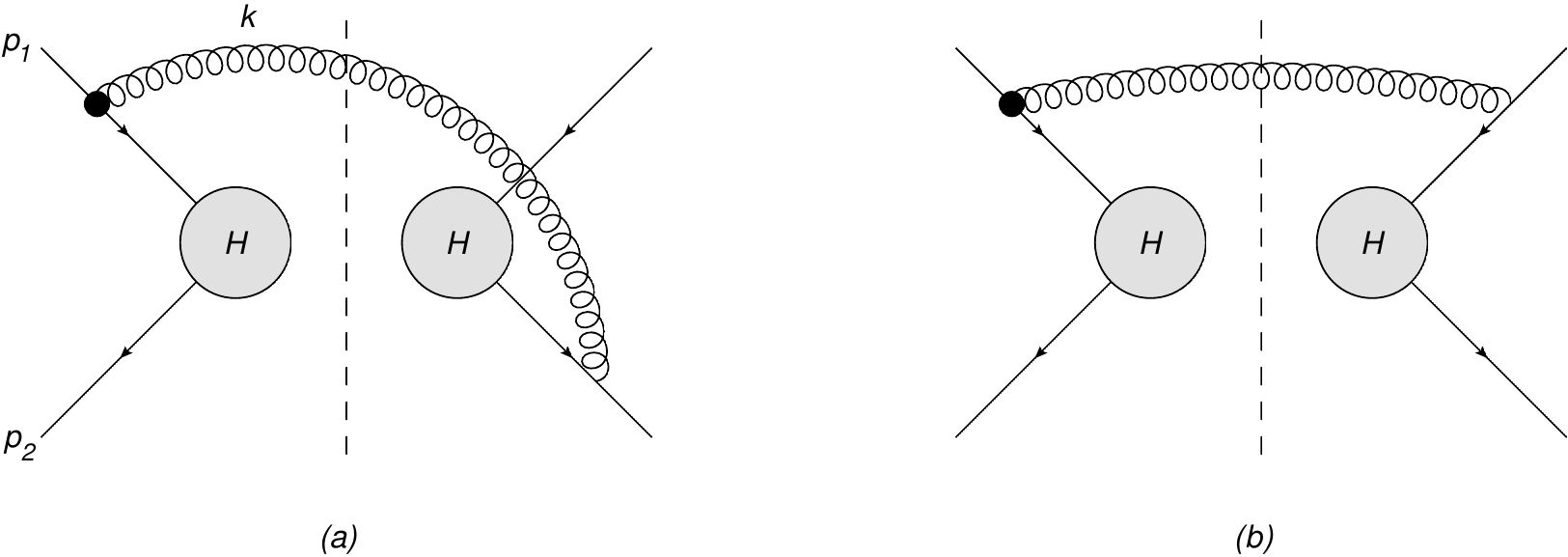}}
\caption{Diagrams contributing to a squared amplitude with a $q \bar{q}$
  initial state, arising from the next-to-soft function $\tilde{S}$ acting on the 
  LO hard interaction ${\cal H}$, as in \eq{SH}. Further diagrams are obtained 
  by reflection about the final state cut, or by interchanging $p_1 \leftrightarrow 
  p_2$.}
\label{fig:NEdiags}
\end{center}
\end{figure}

For the first steps of our derivation, we do not need to specify the final-state
particle content of the process we are studying. Rather, we will consider a 
general hard interaction ${\cal H}$ connecting to an incoming $q \bar{q}$ pair,
such that the LO amplitude is given by \eq{A0def}. Representative diagrams 
contributing to the squared amplitude arising from the next-to-soft function 
at NLO, \eq{SH}, are shown in Figure~\ref{fig:NEdiags}.
We may directly evaluate them using the Feynman rules arising from \eq{fidef}. 
First, we may note that contributions involving $k^2$ vanish, since the radiated 
gluon is on shell. Next, it is convenient to combine the scalar-like and 
spin-dependent emission vertices as
\beq
  \frac{k^\mu}{2 p_i \cdot k} - {\rm i} k_\nu \frac{\Sigma^{\nu \mu}}{p_i \cdot k}
  \, = \, \frac{\slsh{k} \gamma^\mu}{2 p_i \cdot k} \, .
\label{NEvertex}
\eeq
Then, the diagrams of Figure~\ref{fig:NEdiags} yield a NLP contribution
\beqa
  \left| {\cal M} \right|^2_{\rm NLP, \, (a) + (b)} & = & 2 g_s^2 C_F 
  \left(\frac{p_1^\mu}{p_1 \cdot k} - \frac{p_2^\mu}{p_2 \cdot k} \right)
  {\rm Tr} \left[ \slsh{p}_2 {\cal H} \left( \frac{\slsh{k} \gamma^\mu}{2 p_1 \cdot k}
  \right) \slsh{p}_1 {\cal H}^\dag \right] \nonumber \\
  & = & - \, \frac{g_s^2 C_F}{p_1 \cdot k \, p_2 \cdot k} \, {\rm Tr}
  \left[ \slsh{p}_2 {\cal H} \slsh{k} \slsh{p}_2 \slsh{p}_1 {\cal H}^\dag \right] \, ,
\label{NLPcalc1}
\eeqa
where a factor of two for complex conjugate diagrams has already been
included. In order to extract the LO squared amplitude, we may use
an argument similar to one presented recently in Ref.~\cite{DelDuca:2017twk}. 
Writing the decomposition (cf. \eq{Sudakov})
\beq
  \slsh{k} \, = \, \frac{p_2 \cdot k}{p_1\cdot p_2} \slsh{p_1}
  + \frac{p_1 \cdot k}{p_1 \cdot p_2} \slsh{p_2} + \slsh{k}_T \, ,
\label{Sudakov2}
\eeq
and substituting into \eq{NLPcalc1} reveals that the term involving transverse 
momentum occurs linearly in the squared matrix element, leading to an odd 
integrand which vanishes upon integrating over $k_T$. This contribution can 
thus be ignored, leading effectively to the expression
\beq
  \left| {\cal M} \right|^2_{\rm NLP, \, (a) + (b)} \, = \, - \, \frac{2 g_s^2
  C_F}{p_1 \cdot k} \, {\rm Tr} \left[ \slsh{p}_2 {\cal H} \slsh{p}_1 {\cal H}^\dag
  \right] \, .
\label{NLPcalc2}
\eeq 
Notably, in \eq{NLPcalc2} the LO squared matrix element is
factored out.  Combining this with diagrams obtained from those
of Figure~\ref{fig:NEdiags} by interchanging $p_1$with $p_2$, summing
over spins and colours, and dividing out the LO cross section one
easily obtains an expression for the real emission contribution to the
one-loop next-to-soft function. Notice that at NLP singularities as
$z\rightarrow 1$ are integrable: thus, there is no need to combine
real emission with virtual corrections in order to generate LL
contributions, and the NLP soft function reads
\beqa
  {\cal S}_{\rm NLP}^{(1)} \left( z, Q^2, \epsilon \right) & = & 
  - 2 \mu^{2 \epsilon} g_s^2 \, C_F \int \frac{d^d k}{(2 \pi)^{d -1}} \, 
  \delta_+ (k^2) \, \delta \! \left(1 - z - \frac{2 k \cdot (p_1 + p_2)}{\hat{s}} 
  \right) \nn \\ && \hspace{3cm} \times 
  \left(\frac{1}{p_1 \cdot k} + \frac{1}{p_2 \cdot k} \right)  \, . 
\label{NLPint0}
\eeqa
The integration over the real gluon phase space can be carried out 
straightforwardly using the Sudakov decomposition of \eq{Sudakov}, 
and the subsequent change of variables in \eq{kpmtrans}, with the result
\beq
  {\cal S}_{\rm NLP}^{(1)} \left( z, Q^2, \epsilon \right) \, = \, 
  - \frac{2 \as C_F}{\pi} \left(\frac{\bar \mu^2}{Q^2}\right)^{\eps} 
  \frac{e^{\eps \gamma_E}\Gamma^2(- \eps)}{\Gamma(1 - \eps) 
  \Gamma(- 2 \eps)} \,(1 - z)^{ - 2 \eps} \,. 
\label{NLPint1}
\eeq
Taking the Mellin transform we find
\beqa 
  {\cal S}_{\rm NLP}^{(1)} \left( N, Q^2, \epsilon \right) 
  & \equiv & \int_0^1 d z \, z^{N - 1} {\cal S}_{\rm NLP}^{(1)} 
  \left( z, Q^2, \epsilon \right) \nn \\
  & = & - \frac{2 \as C_F}{\pi} \left(\frac{\bar \mu^2}{Q^2}\right)^{\eps} 
  \frac{e^{\eps \gamma_E}\Gamma(- \eps) \Gamma(N)}{\Gamma(1 - 2 \eps + N)}  \nn \\
  & = & \frac{2 \alpha_s C_F}{\pi} \left( \frac{\bar{\mu}^2}{Q^2}
  \right)^{\epsilon} \frac{1}{N} \left[ \frac{1}{\epsilon} + 2 \psi^{(0)} (N + 1) 
  + 2 \gamma_E \right] + {\cal O} (\epsilon) \, .
\label{NLPint20}
\eeqa
The leading behaviour as $N \rightarrow \infty$ is
\beq
  {\cal S}_{\rm NLP}^{(1)} \left( N, Q^2, \epsilon \right) \, = \, 
  \frac{2 \alpha_s C_F}{\pi} \left( \frac{\bar{\mu}^2}{Q^2} \right)^{\epsilon}
  \left[ \, \frac{1}{\epsilon} \, \frac{1}{N} + \frac{2 \log N}{N} + \ldots \right] \, ,
\label{NLPint3}
\eeq 
where the ellipsis denotes terms which are non-singular in $\epsilon$
and non-logarithmic in $N$, as well as terms suppressed by further
powers of $N$. We see that the NLP soft function generates
contributions which are suppressed by (at least) a single power of $N$
compared to LP, as expected. \eq{NLPint3} must now be combined with 
the LP soft function given \eq{eikNLO8}, which itself includes subleading 
terms in $N$ space arising from the Mellin transformation from $z$ space.
The result is
\beq
  {\cal S}_{\rm LP + NLP} \left(N, Q^2, \epsilon \right) \, = \, 
  \frac{2 \alpha_s C_F}{\pi} \left( \frac{\bar{\mu}^2}{Q^2} \right)^{\epsilon}
  \left[ \frac{1}{\epsilon} \left( \log N + \frac{1}{2N} \right)
  + \log^2 N + \frac{\log N}{N} \right] \, .
\label{SLP+NLP}
\eeq
As explained above, we may directly exponentiate \eq{SLP+NLP}, and
combine it with the LO cross-section. Furthermore, the collinear pole can
be absorbed in the quark distributions, for which we again use the
$\overline{\rm MS}$ factorisation scheme. To this end, we generalise 
\eq{qLL} to
\beq
  q_{\rm LL, \, NLP} \! \left( N, Q^2 \right) \, = \, q_N (Q^2)
  \exp \left[ \frac{\alpha_s C_F}{\pi} \, \frac{1}{\epsilon}
  \left( \log N + \frac{1}{2N} \right) \right] \, ,
\label{qNLP}
\eeq 
and similarly for the antiquark. Note that it is important in \eq{qNLP} that 
we correctly kept track of subleading terms in the Mellin transform of the 
LP soft function. The cross-section at NLP in the threshold expansion and
at LL accuracy then becomes 
\beqa
  \left. \int_0^1 d \tau \, \tau^{N - 1} \, \frac{d \sigma_{\rm DY}}{d \tau} 
  \right|_{\rm LL, \, NLP} & = & \sigma_0 (Q^2) \, q_{\rm LL, \, NLP} \!  
  \left( N, Q^2 \right) \, \bar{q}_{\rm LL, \, NLP} \! \left(N, Q^2 \right) 
  \nn \\ & & \quad \times \exp \left[ \frac{2 \alpha_s C_F}{\pi} 
  \left(\log^2 \! N + \frac{\log N}{N} \right) \right] \, .
\label{NEresum}
\eeqa 
Upon expanding the exponential factor in powers of $\alpha_s$, we may
now perform the inverse Mellin transform of the partonic cross-section
order by order, using the results of Appendix~\ref{app:Mellin}, to get
\beq
  \Delta(z, Q^2)|_{\rm LL, \, NLP} \, = \,  \left( \frac{2 \alpha_s C_F}{\pi} \right)^m 
  \!\! \frac{1}{(m - 1)!}  \left[2 \left( \frac{\log^{2 m - 1}(1 - z)}{1 - z} \right)_+ \! 
  - 2 \log^{2 m - 1}(1 - z) \right] \, .
\label{LLNLP2}
\eeq
This is in complete agreement with (and indeed provides an independent proof 
of) the result of Ref.~\cite{Moch:2009hr}, which argued (consistently with previous 
observations~\cite{Kramer:1996iq,Laenen:2008ux}) that the LL NLP terms at any 
order have a coefficient which is always the negative of that of the corresponding 
leading logarithmic plus distribution. The origin of this phenomenon can be traced to 
the coefficient of the $\epsilon$ pole in \eq{NLPint3}. Given that this pole represents 
a collinear singularity that must be absorbed in the parton distributions, it must 
emerge from the NLP contribution to the LO DGLAP splitting kernel that governs 
such terms. More specifically, the collinear poles in the NLO Drell-Yan cross-section 
have the form (see for example Ref.~\cite{Ellis:1991qj})
\beq
  - \frac{2}{\epsilon} \, P_{qq}^{(0)} \, ,
\label{Pqqeps}
\eeq
where the factor of 2 arises from having collinear singularities associated with 
either of the incoming partons. The splitting function can be expanded near 
threshold as
\beq 
  P^{(0)}_{qq} (z) \, = \, \frac{\alpha_s}{2\pi} \, C_F \left[ \frac{2}{(1 - z)_+} - 2 + 
  \ldots \right] \, .
\label{Pqq}
\eeq 
where the second term gives the NLP contribution in $z$-space, whose Mellin 
transform is
\beq
  \int_0^1 dz z^{N - 1} \, \left. P^{(0)}_{qq}(z) \right|_{\rm NLP} \, = \,
  - \frac{\alpha_s C_F}{\pi} \frac{1}{N} \, .
\label{melPqq}
\eeq
We thus expect the collinear pole of the NLP contribution to the next-to-soft 
function in Mellin space to be given by
\beq
  \frac{2 \alpha_s \,C_F}{\pi} \, \frac{1}{N} \, \frac{1}{\epsilon} \, ,
\eeq
which is indeed observed in \eq{NLPint3}. We see that the next-to-soft function 
generates the correct NLP correction to the splitting kernel as expected. This 
in turn dictates the LL behaviour in the finite part: indeed, in $z$-space, this 
contribution arises completely from an overall $\epsilon$-dependent power 
of $(1-z)$, dressing the pole term. Thus, ensuring that the NLP behaviour of 
the pole term is correct is sufficient to describe also the finite part\footnote{This
  story becomes more complicated in $N$-space, as can be seen from
  the fact that \eq{NLPint20} contains a number of contributions that are
  subleading in $N$, all of which can ultimately be traced to a power of $(1-z)$ 
  in $z$-space.}. Note that the fact that all NLP information in the DGLAP 
splitting function is correctly generated by the next-to-soft expansion provides a
test of the statement made earlier, that all LL threshold effects at (N)LP arise 
from radiation that is (next-to) soft, in addition to being collinear. This in
turn confirms that, at LL accuracy, one may neglect radiative jet 
functions~\cite{DelDuca:1990gz,Bonocore:2015esa,Bonocore:2016awd,
Gervais:2017yxv,Gervais:2017zky,Gervais:2017zdb}.

\eq{NEresum} resums the leading-logarithmic behaviour of the Drell-Yan cross
section at LP and NLP in the threshold expansion: it completely agrees with 
expectations from the literature~\cite{Laenen:2008ux,Moch:2009hr}, and thus 
with the recent SCET analysis of Ref.~\cite{Beneke:2018gvs}, which cross-checked 
against the same references. We emphasise that, of course, at leading power 
there is no need to limit the resummation to leading logarithms: this was
done in \eq{NEresum} only for simplicity, and to underline the close connection 
between leading logarithms at LP and NLP. Indeed, because of the link discussed
above between NLP leading logarithms and the DGLAP kernels, it is straightforward
to incorporate our results in the standard LP resummation formalism: it is 
sufficient to include NLP terms in the quark splitting function. This was argued 
to be appropriate in Refs.\cite{Kramer:1996iq,Laenen:2008ux,Moch:2009hr}, 
and, with the mild assumptions discussed in \secn{sec:NLPmatrix}, it is now 
proven. For completeness, we include here the general resummation ansatz 
introduced in Ref.~\cite{Laenen:2008ux}, which implements this change
in the classic threshold resummation formula of~\cite{Sterman:1987aj,
Catani:1989ne,Catani:1990rp}, together with other proposed modifications
that have effects on subleading NLP logarithms. In Mellin space, the result
of Ref.~\cite{Laenen:2008ux} for the Drell-Yan process can be written as
\beqa
  \ln \Big[ \Delta (N, Q^2) \Big] & = & 
  F_{\rm DY} \left[ \alpha_s (Q^2) \right] + 
  \int_0^1 \, dz \, z^{N - 1} \, \Bigg\{ \frac{1}{1 - z} \, 
  D \left[ \as \left( \frac{(1 - z)^2 Q^2}{z} \right) \right] \nonumber \\ & + &
  2 \,\int_{Q^2}^{(1 - z)^2 Q^2/z} \, 
  \frac{d q^2}{q^2} \, P_{qq}^{\rm \, LP+NLP} \Big[ z, \as (q^2) \Big] \Bigg\}_+ \, .
\label{newresDY}
\eeqa
In \eq{newresDY}, $D(\alpha_s)$ is the well-known LP wide-angle soft 
function for the Drell-Yan process, which has been computed up to three 
loops in~\cite{Moch:2005ky,Laenen:2005uz,Idilbi:2005ni,Ravindran:2005vv}; 
$F_{\rm DY} (\alpha_s)$ resums $N$-independent contributions following 
Ref.~\cite{Eynck:2003fn}; $P_{qq}^{\rm \, LP+NLP}(z, \alpha_s)$ is the
soft expansion of the DGLAP splitting function up to NLP, order by order 
in perturbation theory, which was derived in~\cite{Laenen:2008ux} starting 
from the results of Ref.~\cite{Dokshitzer:2005bf}; furthermore, the `plus' 
prescription is defined to apply only to LP contributions, that are singular 
as $z \to 1$. Leading NLP logarithms in \eq{newresDY} are generated by 
the one-loop NLP contribution to $P_{qq}^{\rm \, LP+NLP}$, as discussed
in this Section. Higher-order terms in the NLP splitting function 
will contribute to, but not exhaust, subleading NLP logarithms; indeed,
the shifts in the phase space boundary and in the argument of the coupling, 
proposed in \eq{newresDY}, and corresponding to a NLP-accurate definition 
of the soft scale of the process, also contribute to subleading logarithms at 
NLP. In Ref.~\cite{Laenen:2008ux}, the accuracy of \eq{newresDY} was 
tested by comparing its expansion to NNLO with existing exact results:
as expected from our current discussion, leading NLP logarithms are exactly 
predicted; furthermore, one observes  that next-to-leading NLP logarithms
are predicted very accurately, and they mostly arise from the NLO
contribution to the NLP splitting function. The small discrepancy arising 
at this level of accuracy (NLL at NLP) between the resummation and the 
finite order result is the first footprint of the need to include radiative jet 
functions at NLP.


\subsection{A brief comparison with the SCET approach}
\label{sec:SCET}

In \secn{sec:resummation}, we have achieved the resummation of leading
NLP logarithms (jointly with all LP logarithms) by applying
essentially diagrammatic arguments, based on the previous analysis of
Ref.~\cite{Laenen:2008gt}, summarised here in
Appendix~\ref{app:replica}. Clearly, these diagrammatic arguments are
in turn based on an underlying factorisation~\cite{Bonocore:2015esa,
  Bonocore:2016awd}, but the argument for resummation is greatly
simplified by the diagrammatic exponentiation properties of the
(next-to-)soft function.  Recently, the resummation of these same
contributions has been achieved for the Drell-Yan process also within
an effective field theory approach based on SCET \cite{Beneke:2018gvs}
(see also Ref.~\cite{Moult:2018jjd}). In this section, we briefly
compare our methods with the SCET analysis, whose physics must
ultimately be equivalent.

The SCET approach relies upon a factorisation of the partonic cross section 
$\Delta (z)$, obtained by expanding the Drell-Yan QCD current  into operators 
defined in terms of effective soft and collinear fields, with a different collinear 
sector associated with each external parton. Hard modes of the field contribute 
through short-distance coefficients of SCET operators, which can be obtained 
by matching to full QCD. Under the assumption that Glauber-type modes of 
the gluon field do not contribute to the relevant observable\footnote{The cancellation 
of Glauber gluons for the Drell-Yan cross section at leading twist was proven in 
Refs.~\cite{Collins:1985ue,Collins:1988ig}. For a detailed treatment of Glauber 
effects in an effective-field-theory context, see Ref.~\cite{Rothstein:2016bsq}.}, 
soft and collinear modes can be factorised into (universal) matrix elements, 
which define \emph{collinear} and \emph{soft} functions, in direct correspondence 
with the jet and the soft functions emerging from the diagrammatic approach 
considered in this work. 

Restricting to the terms relevant for the resummation of leading logarithms up 
to NLP, the momentum-space SCET factorisation for the partonic cross-section 
$\Delta$ introduced in \eq{sigmahadN} takes the form
\beq
  \Delta(z) \, = \,  H (Q^2, \mu_h) \, Q \, \bigg[ S_{\rm DY} \big( Q (1-z), \mu_s \big)
  \, - \,\frac{4}{Q} \int \! d \omega \, S_{2 \xi}(Q (1 - z), \omega, \mu_s) \bigg] \, ,
\label{eq:NLPfactevolved2}
\eeq
where $H(Q^2, \mu_h)$ is the hard function, $S_{\rm DY}$ represents the 
leading-power soft function, equivalent to the one defined in \eq{Sdef}, and 
$S_{2\xi}$ represents that part of the NLP soft function that contributes at 
leading logarithmic accuracy, to be compared with LL form of ${\cal S}_{\rm NLP}$ 
in \secn{sec:resummation}. In \eq{eq:NLPfactevolved2} no collinear functions 
appear explicitly, since hard collinear modes contribute only starting at NLL
accuracy. This conclusion is obtained within SCET by an analysis of all possible 
operators contributing at NLP, and confirmed a posteriori, as we will see below.

Each function in the SCET factorisation depends on a characteristic momentum
scale: $\mu_h \sim Q$ for the hard function, and $\mu_s \sim Q(1 - z)$ for the soft 
functions. Independence of physical observables on the factorisation scales yields 
a renormalisation group equation, which can be solved to resum the large 
logarithms. To this end, it is appropriate to evolve the hard and soft functions in 
\eq{eq:NLPfactevolved2} to a common scale, which in Ref.~\cite{Beneke:2018gvs} 
is chosen to be $\mu_c \sim Q \sqrt{1 - z}$.  To LL accuracy, the evolved hard 
and soft functions can be written as~\cite{Beneke:2018gvs}
\beqa
  \left. H (Q^2, \mu_c) \right|_{\rm LL} & = & 
  \exp \big[4 \, E_{\rm LL} ( \mu_h, \mu_c ) \big] \, H (Q^2, \mu_h) \, , \nn \\ 
  \left. S_{2 \xi} (Q (1 - z), \omega, \mu_c) \right|_{\rm LL} & = &
  \frac{2 C_F}{\beta_0} \, \ln \frac{\alpha_s(\mu_c)}{\alpha_s(\mu_s)} \,\,
  \exp \big[ - 4 \, E_{\rm LL} ( \mu_s, \mu_c ) \big]
  \theta(1 - z) \delta(\omega) \, .
\label{eq:RGE}
\eeqa
Here the evolution factor $E_{\rm LL}$ resums the logarithms, and can be 
expressed in terms of the strong coupling evaluated at different scales, as
\beq
  E_{\rm LL} (\nu, \mu) \, = \, - \int \limits_{\alpha_s(\nu)}^{\alpha_s(\mu)} \!
  d \alpha \, \frac{\Gamma_{\rm cusp}(\alpha)}{\beta(\alpha)}
  \int \limits_{\alpha_s(\nu)}^\alpha \frac{d \alpha'}{\beta(\alpha')} 
  \, \stackrel{\rm LL}{=} \, \frac{C_F}{\beta_0^2} 
  \frac{4 \pi}{\alpha_s(\nu)} \left( 1 - \frac{\alpha_s(\nu)}{\alpha_s(\mu)}
  + \ln \frac{\alpha_s(\nu)}{\alpha_s(\mu)} \right) \, ,
\label{SEvolveLL}
\eeq
where we have introduced the QCD cusp anomalous dimension and beta
function. Using the fact that, for Drell-Yan production, $H (Q^2, \mu_h) = 1 
+ {\cal O} (\alpha_s)$, one can compute $\Delta(z)$ in \eq{eq:NLPfactevolved2} 
with all factors evaluated at the scale $\mu_c$. In order to avail oneself 
of the standard collinear factorisation machinery, one must then evolve
both $\Delta(z)$ and the parton distributions to a generic factorisation
scale $\mu$, exploiting the RG invariance of the physical cross-section.
One then finds
\beq
  \Delta^{\rm LL}_{\rm NLP} (z, \mu) \, = \, - \,  \frac{8 C_F}{\beta_0} \, 
  \exp \big[ 4 \, E_{\rm LL} (\mu_h,\mu) - 
  4 \, E_{\rm LL} (\mu_s, \mu) \big] \,   
  \ln \frac{\alpha_s(\mu)}{\alpha_s(\mu_s)} \, .
\label{eq:NLPsfg}
\eeq
The fact that the evolution of parton distributions (dictated by DGLAP splitting
functions) is consistent at LL accuracy with the evolution of the partonic cross 
section to a generic scale provides an independent check of the fact that 
collinear function (absent in \eq{eq:NLPfactevolved2}) cannot contain leading 
NLP logarithms. This is directly analogous to the observation made here in 
\secn{sec:resummation}, where we noted that the effect of including next-to-soft 
radiation led to reproducing the NLP contribution to the DGLAP kernels, testing
our arguments for not including radiative jet functions in the derivation leading 
to \eq{NEresum}.

In order to compare the result in \eq{eq:NLPsfg} with \eq{LLNLP2}, one needs 
to expand the ratios of running couplings in \eq{eq:NLPsfg} and \eq{SEvolveLL} 
in powers of $\alpha_s(\mu)$. When this is done, the NLP term in \eq{eq:NLPsfg} 
reduces to
\beq
  \Delta^{\rm LL}_{\rm NLP} (z, \mu) \, = \, 
  - 4 \, \frac{\alpha_s}{\pi}  C_F \, \exp \left[ - \frac{2 \alpha_s C_F}{\pi}
  \ln^2 \frac{\mu}{\mu_h} \right] \, \exp \left[ \frac{2 \alpha_s C_F}{\pi}
  \ln^2 \frac{\mu}{\mu_s} \right] \ln \frac{\mu_s}{\mu} \, \theta(1 - z) \, .
\label{eq:NLPsummedfinallog}
\eeq
Upon setting the hard and soft scales to their natural values, $\mu_h = Q$
and $\mu_s = Q (1 - z)$, and choosing (as above) a factorisation scale of
$\mu = Q$, we find
\beq
  \Delta^{\rm LL}_{\rm NLP} (z, \mu) \, = \, - 4 \, \frac{\alpha_s}{\pi} C_F
  \exp \left[ \frac{2 \alpha_s C_F}{\pi} \ln^2(1 - z) \right]
  \ln (1 - z) \, \theta (1 - z) \, ,
\label{eq:NLPsummedfinallog2}
\eeq
which is readily seen to be equivalent to \eq{LLNLP2}.


\subsection{Resummation for general quark-initiated colour-singlet production}
\label{sec:generalqq}

In section \ref{sec:resummation} we have seen how to resum the highest power of 
NLP logs, for the specific case of Drell-Yan production. In fact, the result can 
easily be generalised to the production of $N$ colour singlet particles (which 
may be loop-induced at leading order), with a $q\bar{q}$ initial state. Crucial 
to our arguments will be exponentiation of the next-to-soft function in terms 
of webs~\cite{Laenen:2008gt,Laenen:2010uz}, which implies that the
next-to-soft function has the schematic form
\beq
  {\cal S}_{\rm NLP}\, = \, \exp \left[ \sum_i W_{\rm LP}^{(i)} + 
  \sum_j W_{\rm NLP}^{(j)} \right] \, ,
\label{Sform}
\eeq
where the first sum is over leading power webs composed with eikonal Feynman 
rules, and the second sum is over next-to-leading power webs, containing 
eikonal Feynman rules with at most one next-to-eikonal vertex. Next, we may 
note, as was remarked in Refs.~\cite{Laenen:2008gt,Laenen:2010uz}, that if 
we are only interested in NLP terms in the final result for the cross-section, we
do not in fact have to exponentiate the NLP webs: upon expanding \eq{Sform} 
in powers of the coupling, quadratic and higher powers of the NLP term will give 
NNLP contributions and beyond. Thus, we may formally replace \eq{Sform} with 
the equivalent expression (up to NLP level)
\beq
  {\cal S}_{\rm NLP} \, = \, \exp \left[ \sum_i W_{\rm LP}^{(i)} \right]
  \left( 1 + \sum_j W_{\rm NLP}^{(j)} \right) \, .
\label{Sform2}
\eeq
This expression shows us that, in order to generate a contribution to the 
highest power of the NLP logarithm at any given order, we must take the 
leading logarithmic behaviour from the NLP web term, namely the contribution
proportional to 
\beq
  \alpha_s \, \frac{\log N}{N}
\label{sim1}
\eeq
in Mellin space, and dress this with the leading logarithms coming from the
leading-power soft function. Note in particular that the webs $W_{\rm NLP}^{(i)}$
do not contain terms of the type 
\beq
  \alpha_s^p \, \, \frac{\log^{2 p - 1} N}{N} \, ;
\eeq
such terms will arise in the cross section only through the expansion of the 
exponential in \eq{Sform}, precisely through the interference between 
leading-power and next-to-leading power webs. We can see this directly in 
\eq{NEresum} for Drell-Yan production: upon Taylor-expanding in $\alpha_s$, 
the leading logarithm at NLP comes from a single instance of the leading 
NLP log at ${\cal O}(\alpha_s)$, dressed by arbitrary powers of the leading 
logarithm at leading power. 

\begin{figure}
\begin{center}
  \scalebox{0.8}{\includegraphics{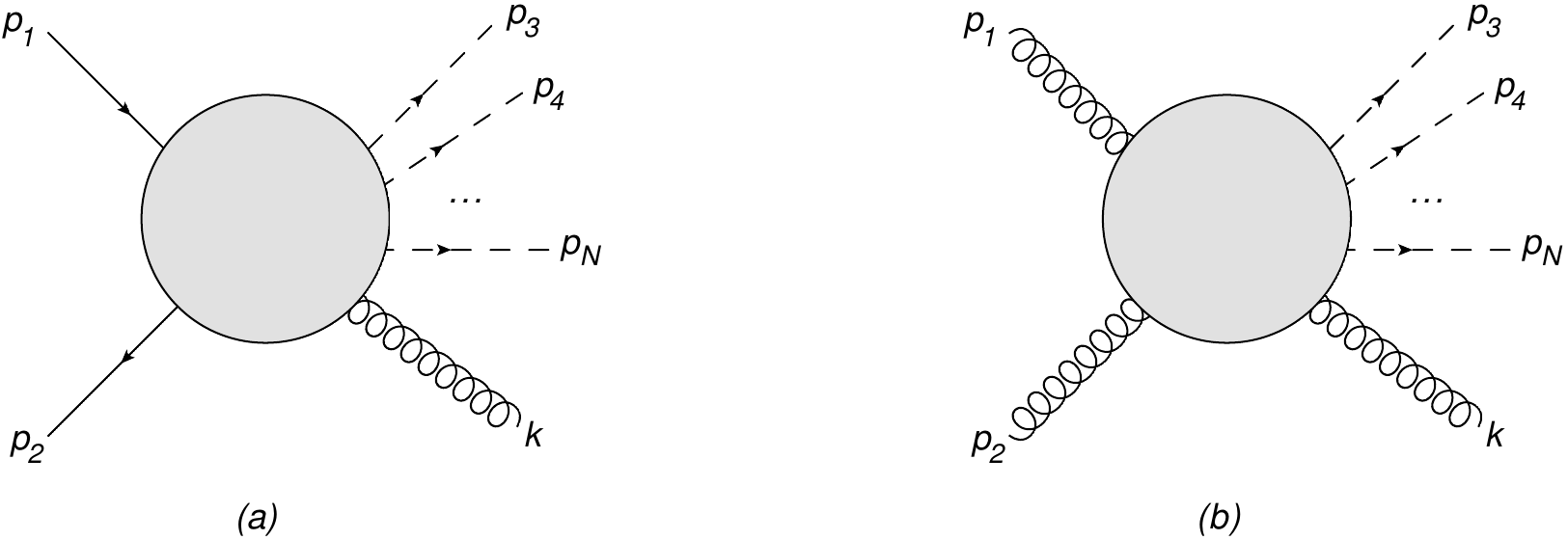}}
  \caption{Production of $N$ colour singlet particles with (a) a
  $q \bar{q}$ initial state; (b) a $gg$ initial state.}
\label{fig:qqggfig}
\end{center}
\end{figure}

For arbitrary processes, we must broaden the discussion presented for the 
Drell-Yan case to include an additional next-to-soft contribution, associated 
with the orbital angular momentum of incoming particles, which combines
with the spin angular momentum present in the next-to-soft function to build 
a gauge-invariant result.  To this end, let us consider the effect of a single 
emission from the non-radiative amplitude; this has been examined recently 
in Ref.~\cite{DelDuca:2017twk}, and we will now present a short summary 
of that discussion, before drawing consequences for the present paper. 
We label momenta as shown in Figure~\ref{fig:qqggfig}(a), 
and we write the LO non-radiative amplitude for a $q \bar{q}$-initiated 
process as
\beq
  {\cal M}^{(q \bar{q})}_{\rm LO} \big( \{p_i\} \big) \, = \, \bar{v}(p_2) \,
  M^{(q \bar{q})}_{\rm LO} \big( \{p_i\} \big) \, u(p_1) 
  \, = \, \bar{v}(p_2) \,
  {\cal H}^{(q \bar{q})}_{\rm LO} \big( \{p_i\} \big) \, u(p_1)\, .
\label{Adef}
\eeq
where $\{p_i\}$ are the incoming parton momenta, and ${\cal H}_{\rm LO}$ 
is the LO hard function, which coincides with the LO stripped matrix element 
$M^{(q \bar{q})}_{\rm LO}$. Let us now consider the radiative amplitude with 
external wave functions removed, which we denote by $M^{(q \bar{q} g)}_\sigma$.
As shown in Ref.~\cite{DelDuca:2017twk}, this amplitude, up to NLP order, can 
be decomposed as
\beq
  M^{(q \bar{q} g) \, \sigma}_{\rm NLP} \, = \, 
  M^{(q \bar{q} g) \, \sigma}_{\rm scal.} + 
  M^{(q \bar{q} g) \, \sigma}_{\rm spin} + 
  M^{(q\bar{q} g) \, \sigma}_{\rm orb.} \, ,
\label{Asigmadef}
\eeq
where the first (second) term on the right-hand side originates from
the spin-independent (spin-dependent) part of the next-to-soft function, 
while the third term corresponds to the orbital  angular momentum 
contribution discussed above. The squared real emission amplitude, 
summed over colours and spins, is then given by~\footnote{Note that 
\eq{Asq}, as written, contains terms at NNLP, arising from squaring 
NLP contributions. Such terms should be neglected in the final result, 
given that accuracy is guaranteed up to NLP only.}
\beq
  \left|{\cal M}^{(q \bar{q} g)}_{\rm NLP} \big( p_1, p_2, k \big) \right|^2 \, = \,
  - \, \sum_{\rm colours}  {\rm Tr} \left[\slsh{p}_1 \, 
  M^{(q \bar{q} g) \, \sigma}_{\rm NLP} \slsh{p}_2 \, M^{(q \bar{q} g) \, 
  \dag}_{\rm NLP , \, \sigma} \right] \, ,
\label{Asq}
\eeq
where we have used the gluon polarisation sum
\beq
  \sum_{\lambda} \varepsilon^{(\lambda)}_\sigma (k) \,
  \varepsilon^{(\lambda) *}_\tau (k) \, = \, - \eta_{\sigma\tau} \, ,
\label{Pcaldef2}
\eeq
since the contribution of unphysical polarisations vanishes when just a 
single gluon is radiated. The various contributions to \eq{Asigmadef} have 
been calculated explicitly in Ref.~\cite{DelDuca:2017twk}, and one obtains
for squared matrix element, summed over colours and spins, the expression
\beq
  \left| {\cal M}^{(q \bar{q} g)}_{\rm NLP} (p_1, p_2, k) \right|^2
  \, = \, g_s^2 C_F \frac{\hat{s}}{p_1\cdot k \, p_2\cdot k}
  \left| {\cal M}^{(q \bar{q})}_{\rm LO} \big(p_1 + \delta p_1, 
  p_2 + \delta p_2 \big) \right|^2 \, ,
\label{ANLPsq}
\eeq
where initial state momenta in the LO matrix element have been shifted 
according to
\beq
  \delta p_1 \, = \, - \frac12 \left(\frac{p_2\cdot k}{p_1\cdot p_2} p_1^\alpha
  - \frac{p_1 \cdot k}{p_1 \cdot p_2} p_2^\alpha + k^\alpha\right) \, , \quad
  \delta p_2 \, = \, - \frac12 \left(\frac{p_1\cdot k}{p_1\cdot p_2} p_2^\alpha
  - \frac{p_2\cdot k}{p_1\cdot p_2} p_1^\alpha + k^\alpha \right) \, .
\label{momshifts}
\eeq
In words, the NLP squared amplitude for single real emission (summed over 
colours and spins) consists of an overall eikonal factor dressing the LO squared 
amplitude, whose incoming momenta are shifted according to \eq{momshifts}. 
These shifts have the effect of rescaling the partonic Mandelstam invariant 
$\hat{s}$ according to
\beq
  \hat{s} \, \rightarrow \, z \hat{s} \, ,
\label{sshift}
\eeq
where the threshold variable $z$ is defined by
\beq
  z \, = \, \frac{P^2}{\hat{s}} \, , \quad P^\mu \, = \, \sum_{i = 3}^{N + 2} p_i^\mu \, ,
\label{zNdef}
\eeq
satisfying the momentum conservation condition 
\beq 
  p_1^\mu + p_2^\mu \, = \, P^\mu + k^\mu \, . 
\label{momconsrad}
\eeq 
Crucially for what follows, all NLP effects in the matrix element are absorbed 
in the momentum shift, so that the prefactor in \eq{ANLPsq} simply dresses 
the shifted matrix element with a leading-power soft emission. We may obtain 
the partonic cross-section for the single real emission contribution by integrating 
over phase space and including flux and spin/colour averaging factors. The
phase space for the $(N+1)$-body final state, with momenta labelled as
in Figure~\ref{fig:qqggfig}(a), may be written in factorised form as
\beq 
  \int d \Phi_{N + 1} \left(P + k; p_3, \ldots p_{N + 2}, k \right) \, = \,  
  \int \frac{d P^2}{2 \pi} \, d \Phi_2 \left(P + k; P, k \right) \, d \Phi_N
  \left(P; p_3, \ldots p_{N + 2} \right) \, ,
\label{phasespacefac}
\eeq 
namely as the convolution of a two-body phase space for the gluon momentum 
$k$ and the total momentum $P$ carried by colour singlet particles, with the
subsequent decay of the latter into the individual colour singlet momenta
$\{p_i\}$. Parametrising momenta according to
\begin{align}
  p_1 \, =& \, \frac{\sqrt{\hat{s}}}{2} \left(1, 0, \ldots, 0, 1 \right) \, , \quad
  p_2 \, = \, \frac{\sqrt{\hat{s}}}{2} \left(1, 0, \ldots, 0, - 1 \right) \, , 
  \notag \\
 &\quad k \, = \, \frac{(1 - z) \sqrt{\hat{s}}}{2} \left(1, 0, \ldots, \sin \chi, \cos \chi \right) \, ,
\label{momparamN}
\end{align}
\eq{phasespacefac} becomes~\cite{DelDuca:2017twk} 
\beq
  \int d \Phi_{N+1} \left(P + k; \{ p_i\}, k \right) \, = \, 
  \frac{1}{16 \pi^2 \, \Gamma(1 - \e)} \left( \frac{4 \pi}{\hat{s}} \right)^\e
  \int d P^2 \, d \Phi_N^{(z)} \, d y \, (1 - z)^{1 - 2 \e} \, \Big[ y (1 - y) \Big]^{- \e} \, ,
\label{dPS2}
\eeq
with
\beq
  y \, = \, \frac{1 + \cos \chi}{2} \, ,
\label{ydef}
\eeq 
and where $d\Phi_N^{(z)}$ denotes the phase space for the $N$ colour singlet 
particles, but with kinematics shifted according to \eq{sshift}. Then, the partonic 
cross-section including a single additional emission (up to NLP level) takes 
the form
\beq
  \widehat{\Delta}^{(q \bar{q})}_{\rm \, NLP} (z, \epsilon) \, = \, K_{\rm NLP} (z,\epsilon)
  \, \hat{\sigma}_{\rm LO}^{(q \bar{q})} \big( z \hat{s} \big) \, ,
\label{partxsec}
\eeq
where 
\beqa
  K_{\rm NLP} \left( z, \e \right) & = & \frac{\alpha_s}{\pi} \, C_F
  \left( \frac{4 \pi \mu^2}{\hat{s}} \right)^\e
  z \, (1 - z)^{- 1 - 2 \e} \frac{\Gamma^2 (- \e)}{\Gamma(- 2 \e)
  \Gamma(1 - \e)} \, ,
\label{dsigmadz2}
\eeqa 
while the LO cross-section with shifted kinematics is given by
\beq
  \hat \sigma^{(q \bar{q})}_{\rm LO} \left( z \hat{s} \right) \, = \, 
  \frac{1}{2 (z \hat{s})}\frac{1}{4 N_c^2} \, \int d \Phi_N^{(z)} \,
  \Big| M^{(q\bar{q})}_{\rm LO} \big( p_1 + \delta p_1, p_2 + \delta p_2 
  \big) \Big|^2 \, .
\label{sigma0def}
\eeq
As discussed above, the generalisation of \eq{partxsec} to all orders is obtained 
by dressing the single-emission cross-section with a further arbitrary number 
of leading-power soft gluon emissions. In \eq{ANLPsq}, this has the effect 
of replacing the prefactor -- whose form is obtained from the soft function 
at ${\cal O}(\alpha_s)$ -- with that obtained from the all-order leading-power 
soft function. Furthermore, the $(N+m)$-body phase space for the emission of 
$N$ colour singlet particles and $m$ additional gluons, with momenta $\{p_i\}$ 
and $\{k_j\}$ respectively, factorises as in \eq{dPS2}, and one may write
\beq
  \int d \Phi_{N + m} \bigg(P + \sum_{j=1}^m k_j; \{p_i\}, \{k_j\} \bigg) \, = \, 
  \int \frac{d P^2}{2 \pi} \, d \Phi_{m+1} \bigg(P + \sum_{j=1}^m k_j; P, \{k_j\} 
  \bigg) \, d \Phi_N \big(P; \{p_i\} \big) \, ,
\label{dPSm}
\eeq
so that \eq{partxsec} can be straightforwardly replaced with
\beq
  \widehat{\Delta}_{\rm NLP}^{(q \bar{q})}(z, \epsilon) \, = \, z \, 
  {\cal S}_{\rm LP} (z, \epsilon) \, 
  \hat{\sigma}_{\rm LO}^{(q\bar{q})} \big( z \hat{s} \big) \, ,
\label{partxsecresum}
\eeq
where the factor of $z$ on the right-hand side originates from having shifted 
the flux factor in \eq{sigma0def}. In \eq{partxsecresum}, ${\cal S}_{\rm LP}
(z, \epsilon)$ is the leading-power soft function, defined to include integration 
over the soft gluon phase space, as in \eq{Stildedef}: it will contain residual 
collinear poles in $\epsilon$ that must be absorbed into the quark distribution 
functions, as was done in \eq{qNLP}. We may then resum leading-logarithmic
LP and NLP terms in the partonic cross-section as follows. First, we notice 
that the leading-order partonic cross section with shifted kinematics becomes 
a function of $z \hat s = Q^2$, which is the physically measured invariant mass 
that must be kept fixed: we can therefore treat the factor $\hat{\sigma}_{\rm 
LO}^{(q \bar{q})}$ as independent of $z$, writing
\beq
  \widehat{\Delta}_{\rm \, NLP}^{(q \bar{q})}(z, \epsilon) \, = \, z \, 
  {\cal S}_{\rm LP} (z, \epsilon) \, 
  \hat{\sigma}_{\rm LO}^{(q \bar{q})}(Q^2) \, ,
\label{qqresum1}
\eeq
Taking the Mellin transform we find
\beq
  \int_0^1 dz z^{N - 1} \widehat{\Delta}_{\rm \, NLP}^{(q \bar{q})} (z, \epsilon)
  \, = \, {\cal S}_{\rm LP} (N + 1, \epsilon) \, 
  \hat\sigma_{\rm LO}^{(q \bar{q})} (Q^2) \, .
\label{qqresum2}
\eeq
Since the leading-power soft function is insensitive to the details of the hard process, 
we can directly use \eq{eikNLO8} for the soft factor. Removing collinear poles,
using exponentiation, and keeping track of NLP terms that arise from the Mellin 
transform of the LP soft function, we find
\beq
  \int_0^1 dz z^{N - 1} \Delta_{\rm \, NLP}^{(q \bar{q})}(z) \, = \, 
  \hat \sigma^{(q\bar{q})}_{\rm LO}(Q^2) 
\exp\left[ \frac{2\alpha_s C_F}{\pi}\log^2(N) \right] 
\left(1+\frac{2\alpha_s C_F}{\pi} \frac{\log N}{N}\right).
\label{qqresum3}
\eeq
This simple result resums leading logarithmic terms in Mellin space at both LP 
and NLP, in the partonic cross-section, for a general quark-induced colour
singlet production process. In the Drell-Yan case, it agrees with \eq{NEresum}, 
thus providing an important cross-check of \eq{qqresum3}. As was the case 
for the Drell-Yan process, \eq{qqresum3} can be generalised to include the 
complete known result for the resummation of leading-power subleading 
logarithms, yielding an expression identical to \eq{newresDY} for the 
resummed partonic cross section $\Delta(N,Q^2)$. Indeed, the orbital angular 
momentum contribution that was trivial for the Drell-Yan cross section, due to 
the point-like nature of the Born process, will result in a shift of the center-of-mass 
energy $\hat{s}$, which must be applied to the Born cross section, with consequences 
that will depend on the particular process and observable being considered
(it would for example be non-trivial for loop-induced processes). The Sudakov
exponent, on the other hand, will be unaffected, so that \eq{newresDY} will 
still apply.

In this section, we have seen that resummation of LL terms is possible at both 
LP and NLP for the general production of $N$ colour-singlet particles, in the 
$q\bar{q}$ channel. Similar arguments may be made for gluon-initiated 
processes, as we discuss in the following section.


\subsection{Resummation for general gluon-initiated colour-singlet production}
\label{sec:generalgg}

\begin{figure}
\begin{center}
  \scalebox{0.5}{\includegraphics{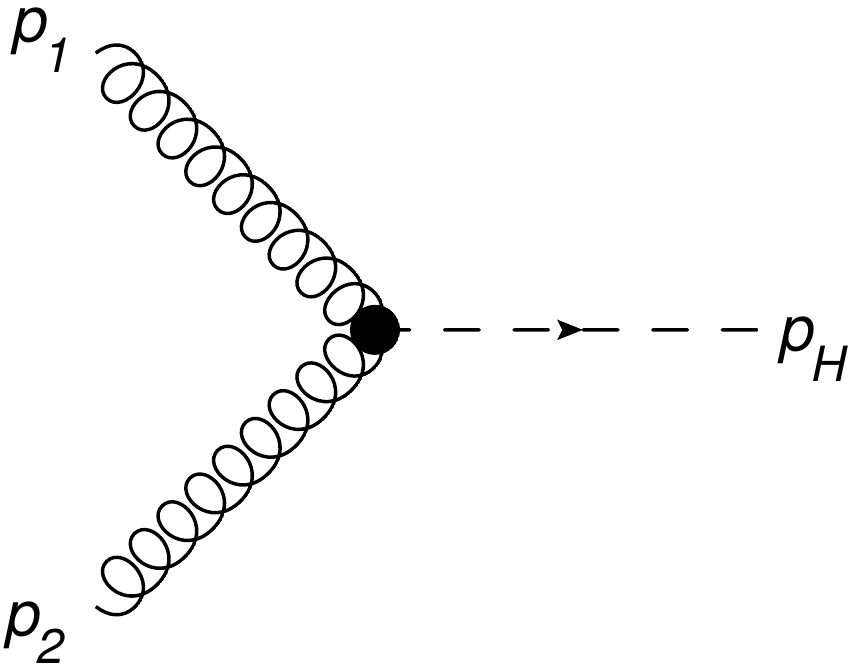}}
  \caption{Higgs boson production via gluon-gluon fusion, where
  $\bullet$ denotes the effective coupling resulting from the integration of the 
  top quark loop. }
\label{fig:Higgs}
\end{center}
\end{figure}

In \secn{sec:generalqq} we considered the production of a generic colour singlet
final state in quark-antiquark scattering. A similar analysis can be made for 
gluon-initiated processes: one may obtain leading logarithmic NLP contributions 
by combining the next-to-soft function with orbital angular momentum contributions. 
As for the quark case of \secn{sec:generalqq}, we can then dress the effect of a 
single gluon emission at NLP with an arbitrary number of leading-power soft 
gluon emissions. The case of single emission has been studied alongside the 
quark case in Ref.~\cite{DelDuca:2017twk}, leading to a result identical in form 
to \eq{ANLPsq} for the squared amplitude. Indeed one finds
\beq 
  \left| {\cal M}^{(ggg)}_{\rm NLP} \left(p_1, p_2, k \right) \right|^2 \, = \, 
  g_s^2 C_A \frac{\hat{s}}{p_1 \cdot k \, p_2 \cdot k} \left| {\cal M}^{(gg)}_{\rm LO} 
  \left(p_1 + \delta p_1, p_2 + \delta p_2 \right) \right|^2 \, .
\label{ANLPshift}
\eeq
As in the quark case, this takes the form of the LO non-radiative transition
probability, with kinematics shifted according to \eq{momshifts}, dressed by 
a single leading-power soft emission, whose colour factor in this case 
reflects the emission from an initial-state gluon rather than an initial-state 
(anti)-quark. The factorisation of phase space will be identical to the previous
section, given that this is independent of the particle species. One then obtains 
the resummed result
\beq
  \widehat{\Delta}_{\rm \, NLP}^{(gg)} (z, \epsilon) \, = \, z \, {\cal S}_{\rm LP} 
  (z, \epsilon) \, \hat{\sigma}_{\rm LO}^{(gg)}(z \hat s) \, ,
\label{ggxsecresum}
\eeq
where the soft function on the right-hand side is defined in terms of Wilson lines 
in the adjoint representation. One may then follow similar arguments to those 
leading to \eq{qqresum3}, yielding
\beq
  \int_0^1 dz z^{N - 1} \Delta_{\rm \, NLP}^{(gg)} (z) \, = \, 
  \hat \sigma^{(gg)}_{\rm LO} (Q^2) \exp \left\{ \frac{2\alpha_s C_A}{\pi}
  \log^2(N) \right\} \left( 1 + \frac{2 \alpha_s C_F}{\pi} \frac{\log N}{N} \right) \, . 
\label{ggresum}
\eeq
A check of these results is that it reproduces known LP, and conjectured
NLP results for Higgs boson production, in the large top mass limit. As is 
well-known, the LO process consists of an effective coupling between the 
Higgs boson and a pair of gluons, as shown in Figure~\ref{fig:Higgs}. 
Higher-order contributions near threshold have been discussed for example 
in Ref.~\cite{deFlorian:2014vta}, which expressed the hadronic cross section 
for the $gg$ channel as
\beqa
  \sigma_H \big(s, m_H^2 \big) & = & \tau \tilde{\sigma}_0 \, 
  \int_0^1 \frac{d x_1}{x_1} \int_0^1\frac{d x_2}{x_2} \, g(x_1,\mu^2) \, 
  g(x_2, \mu^2) \nn \\ && \hspace{1cm} \times \, \int_0^1 d z \, 
  \delta \! \left( z - \frac{\tau}{x_1 x_2} \right) 
  c_{gg} \! \left( z, \alpha_S(\mu^2), \frac{m_H^2}{\mu^2} \right) \, .
\label{sigmaH}
\eeqa
Here $g (x_i, \mu^2)$ is the gluon distribution, we have set the factorisation 
and the renormalisation scales to the common value $\mu$, and $c_{gg}$ a
perturbative coefficient function. Furthermore, we have introduced the quantities
\beq
  \tilde{\sigma}_0 \, = \, \frac{\pi C^2(\mu^2)}{64 v^2} \, , \qquad
  C (\mu^2) \, = \, - \frac{\alpha_s}{3 \pi} 
  \left( 1 + 11 \, \frac{\alpha_s(\mu^2)}{4\pi} + {\cal O} 
  \left( \alpha_s^2 \right) \right) \, ,
\label{sigma0defH}
\eeq
which normalises the LO cross-section, and where $v$ is the Higgs field 
vacuum expectation value. With the normalisation adopted in \eq{sigmahad} 
we have 
\beq
  c_{gg} \, = \, \frac{\Delta^{(gg)}_{\rm \, NLP} (z)}{z \, \tilde \sigma_0} \, = \, 
  {\cal S}_{\rm LP, fin.} (z) \, \frac{\hat{\sigma}_{\rm LO}^{(gg)} 
  (z \hat s)}{\tilde \sigma_0} \, ,
\label{partgg}
\eeq
where collinear poles in ${\cal S}_{\rm LP} (z, \epsilon)$ have already 
been factorised into the gluon distributions, leaving a finite remainder 
${\cal S}_{\rm LP, fin.} (z)$. We may now use (see for example 
Eqs.~(5.1) and (5.4) in Ref.~\cite{DelDuca:2017twk}) the fact that 
\beq
  \hat{\sigma}_{\rm LO}^{(gg)} (z \hat{s}) \, = \, z \, \tilde{\sigma}_0 \, ,
\label{sigmazH}
\eeq
to identify 
\beq
  \left. c_{gg} \right|_{\rm LL} \, = \, \big[ 1 - (1 - z) \big] \, 
  {\cal S}_{\rm LP, fin.} (z) + {\cal O} (1 - z) \, .
\label{cggres}
\eeq
This in turn implies that the coefficient of the LL NLP term in $c_{gg}$ at 
a given order in $\alpha_s$ is related to the LL LP term by a minus sign. 
Furthermore, both sets of terms are related to their counterparts in Drell-Yan 
production by the simple replacement $C_F\rightarrow C_A$, given that the 
LP soft functions in both cases obey `Casimir scaling' to the relevant order.
We thus reproduce the results of Ref.~\cite{deFlorian:2014vta} for the 
resummation of LL NLP logarithms in single Higgs production in the large 
top mass limit. We stress, however, that the results of this section are more 
general: they apply also away from the large top mass limit, although 
\eq{sigmazH} will not apply to processes with a more intricate LO cross 
section, and it will be necessary to consider the more general expression 
in \eq{ggresum}. Also for gluon-initiated processes, subleading LP logarithms
can be included, and the result will take the general form of \eq{newresDY}:
in this case, the gluon DGLAP splitting functions will be involved, while 
the soft function for gluon annihilation can be obtained from the quark 
case by Casimir scaling, at least up to three loops.


\section{Conclusion}
\label{sec:discuss}

In this paper, we have developed a formalism for resumming leading-logarithmic 
(LL) threshold contributions to perturbative hadronic cross-sections, at next-to-leading 
power (NLP) in the threshold variable. This generalises previous approaches at 
leading power (see for example Refs.~\cite{Sterman:1987aj,Catani:1989ne,
Catani:1990rp,Korchemsky:1993xv,Korchemsky:1993uz,Forte:2002ni,
Contopanagos:1997nh,Becher:2006nr,Schwartz:2007ib,Bauer:2008dt,Chiu:2009mg}),
and applies to the production of an arbitrary colour-singlet final state at LO. 
Our method builds upon the previous work of Refs.~\cite{Laenen:2008gt,
Laenen:2010uz} (and subsequent studies~\cite{Bonocore:2015esa,
Bonocore:2016awd}), which describes leading NLP effects in terms of 
a {\it next-to-soft function}, which can be shown to exponentiate at the 
diagram level, so that the logarithm of the next-to-soft function can be directly
expressed in terms of Feynman diagrams dubbed {\it next-to-soft webs}. In general 
processes, the next-to-soft function must then be supplemented by terms 
involving derivatives acting on the non-radiative amplitude, which can be 
interpreted in terms of the orbital angular momentum of the colliding partons. 
Leading-logarithmic accuracy can then be achieved by dressing the effect 
of a single emission, computed up to NLP level, with the LP soft function. 
In this sense, our results provide a non-trivial generalisation of the so-called {\it
next-to-soft theorems}~\cite{Casali:2014xpa,Cachazo:2014fwa}, which
have recently been intensively studied in a more formal context, for both 
gauge theories and gravity.

We have explicitly reproduced previously conjectured results for both
Drell-Yan production~\cite{Moch:2009hr} and Higgs boson production in
the large top mass limit~\cite{deFlorian:2014vta}. In particular, we
have verified the observation that the LL NLP contribution at a given
order in perturbation theory is generated by including a subleading
term in the DGLAP kernels that accompany the leading pole in
$\epsilon$ in the unsubtracted cross-section. Our reasoning provides a
proof of one of the ingredients building up the resummation ansatz
proposed in Ref.~\cite{Laenen:2008ux}, which was partly based on the
idea of exponentiating NLP contributions to DGLAP splitting
functions. We note again that it is natural, in this context, to
exponentiate NLP contributions to the splitting functions also beyond
leading order in perturbation theory: this step is strongly suggested
by the arguments in Ref.~\cite{Dokshitzer:2005bf}, which were, in
turn, based on the idea of reciprocity between time-like and
space-like splitting kernels.  Ref.~\cite{Laenen:2008ux} verified that
the inclusion in the Sudakov exponent of NLP terms in the NLO DGLAP
kernel is responsible for the bulk of next-to-leading logarithms at
NLP in the Drell-Yan and DIS cross sections.  On the other hand, it is
clear that, beyond leading NLP logarithms, hard collinear effects and
phase space corrections become relevant, and a full resummation can
only be achieved by including in the initial factorisation the
contributions of radiative jet functions, as done for example in
Refs.~\cite{Bonocore:2015esa, Bonocore:2016awd}. For the Drell-Yan
cross section, we have also compared our results with a recent
analysis based on Soft-Collinear Effective Theory
techniques~\cite{Beneke:2018gvs} (see also ref.~\cite{Moult:2018jjd}),
finding complete agreement.

There are many directions for further work. First, of course, is the extension 
of the present results to subleading logarithmic accuracy at NLP. This will 
require a proper treatment of non-factorising phase-space effects for real 
emission contributions, and a thorough study of the radiative jet functions 
introduced in~\cite{DelDuca:1990gz,Bonocore:2015esa,Bonocore:2016awd,
Gervais:2017yxv,Gervais:2017zky,Gervais:2017zdb}. The latter have yet to 
be fully classified in QCD, while considerable progress was recently achieved 
in SCET~\cite{Moult:2019mog}. We note that the quark radiative jet function
needed for quark annihilation processes into electroweak final states is 
currently known to one-loop order~\cite{Bonocore:2015esa,Bonocore:2016awd}, 
which will constitute a key ingredient to extend the present work to subleading 
NLP logarithms. A second direction for further studies is the inclusion of 
processes with final state partons at Born level: in these cases, additional 
threshold contributions associated with hard collinear real radiation are 
expected, as happens at leading power in the threshold variable. An 
analysis of processes of this kind was performed very recently in 
Refs.~\cite{vanBeekveld:2019prq,vanBeekveld:2019cks}. When 
more than one parton is present in the final state, further complications 
due to non-trivial colour flow will have to be handled, as was the case 
at leading power.

In order to move towards phenomenological applications of this formalism, 
another required step will be the inclusion of threshold contributions arising
beyond leading order from different partonic channels, that are not available 
at Born level. For example, for the Drell-Yan process, the quark-gluon channel 
enters at NLO, and it generates Sudakov logarithms suppressed by an overall 
power of the threshold variable, because of the required radiation of a final state 
fermion. The inclusion of such contributions is necessary for consistent treatment 
of (resummed) NLP threshold effects. Another important issue that will need to 
be studied in detail, in order to gauge the impact of NLP resummation on 
phenomenology, is related to exponentiation: as pointed out in this paper, 
when including NLP corrections, exponentiation has to be understood in a 
limited sense, since NLP terms in the Sudakov exponent will generate a large 
set of potentially spurious contributions at NNLP and beyond upon expanding 
the exponential to any finite order. A precise way to
limit the resummation to relevant and well-understood contributions must
therefore be devised, for example by expanding the NLP part of the
Sudakov exponent to fixed order, as was done in this paper. This issue
is closely related to that of matching the resummation to finite order results,
which is likely to be particularly relevant at NLP.

Once these issues are understood, NLP resummation will provide a new
versatile tool to gauge the impact of high-order corrections for a
range of highly topical Standard Model and BSM processes at the Large
Hadron Collider and beyond, significantly enhancing our mastery of
precision high-energy phenomenology.


\section*{Acknowledgments}

This article is based upon work from COST Action CA16201 PARTICLEFACE
supported by COST (European Cooperation in Science and
Technology). JSD and LV are supported by the D-ITP consortium, a
program of NWO funded by the Dutch Ministry of Education, Culture and
Science (OCW). CDW was supported by the Science and Technology
Facilities Council (STFC) Consolidated Grant ST/P000754/1 ``String
theory, gauge theory \& duality'', and by the European Union’s Horizon
2020 research and innovation programme under the Marie
Sk\l{}odowska-Curie grant agreement No. 764850 ``SAGEX''.


\appendix


\section{Exponentiation via the replica trick}
\label{app:replica}

\begin{figure}
\begin{center}
\scalebox{0.6}{\includegraphics{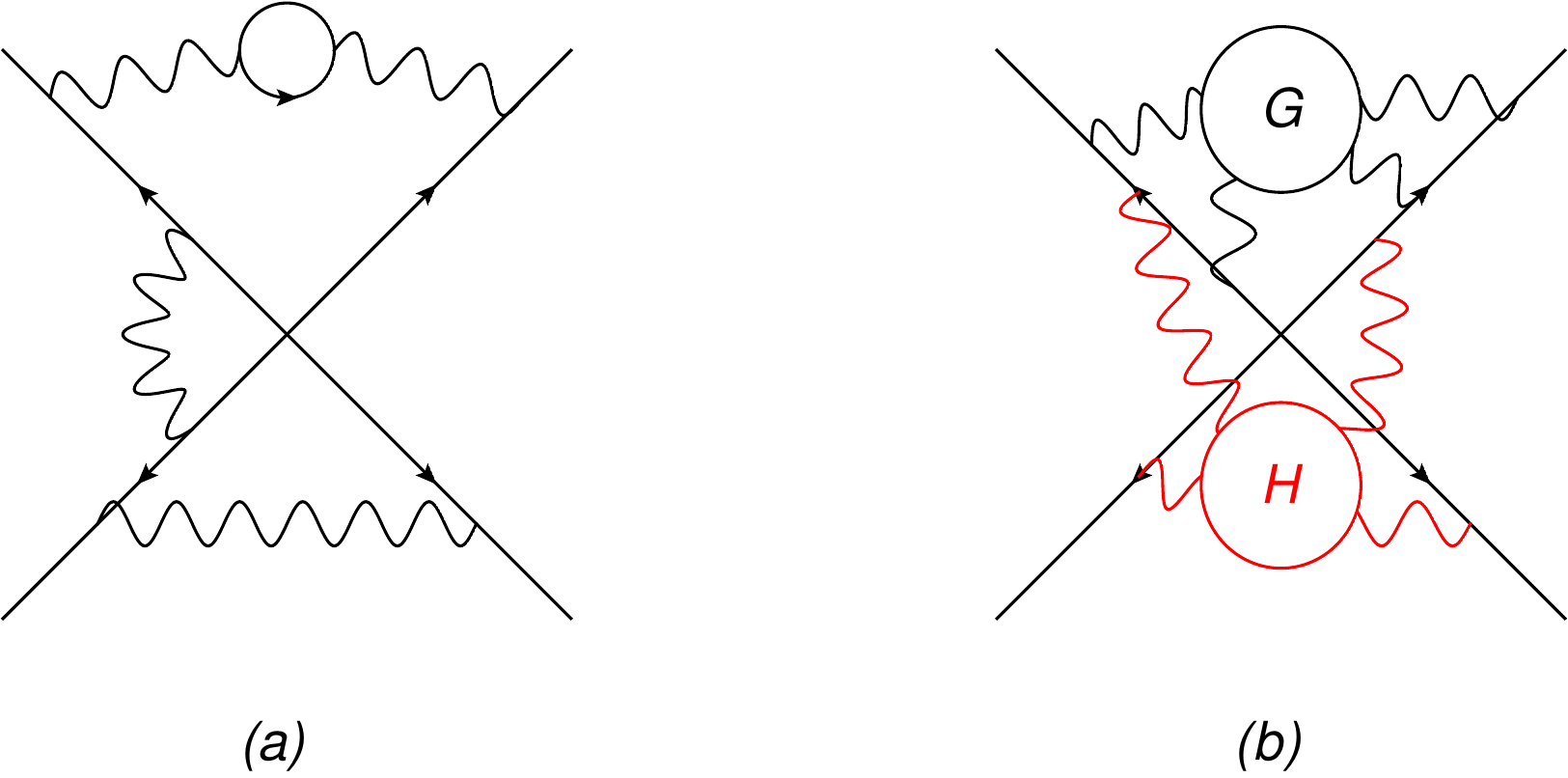}}
  \caption{(a) Example diagram generated by the path integral in \eq{pathintegral}, 
  with straight semi-infinite Wilson lines for illustration, containing three 
  subdiagrams. (b) Example diagram in the replicated theory, with different 
  colours denoting different replicas.}
\label{fig:replicafig}
\end{center}
\end{figure}

In this appendix, we review the methods of Ref.~\cite{Laenen:2008gt}, that 
provide a convenient shortcut for proving that the soft function exponentiates
at the diagrammatic level.  For simplicity, let us first focus on QED rather than
QCD, and consider a single vacuum expectation value of $n$ Wilson line
operators, as would be appropriate for contributions to the soft function 
involving virtual radiation. For the purposes of the present argument, we 
do not need to specify the path of the Wilson lines, and we write
\beq
  {\cal S}_n \, = \, \left\langle 0 \left| \prod_{i = 1}^n \Phi_{i} 
  \right| 0 \right\rangle \, ,
\label{WilsonVEV}
\eeq
where
\beq
  \Phi_i \, = \, \exp \left[ {\rm i} e \!  \int d x_i^\mu A_\mu(x_i) \right] \, .
\label{Phii}
\eeq
In path-integral language, this matrix element may be written as
\beqa
  {\cal S} _n & = & \int {\cal D} A_\mu \left( \prod_{i = 1}^n \Phi_i \right)
  {\rm e}^{ {\rm i} S \left(A_\mu, \bar{\psi}, \psi \right)} \nn \\
  & = & \int {\cal D} A_\mu \, \exp \left[ \sum_{i = 1}^n {\rm i} e \!
  \int d x_i^\mu A_\mu(x_i) + {\rm i} S \left(A_\mu, \bar{\psi}, \psi \right) \right] \, .
\label{pathintegral}
\eeqa
where $S \left(A^\mu, \bar{\psi}, \psi \right)$ is the QED action. Carrying out 
the path integral generates Feynman diagrams in which multiple Wilson lines 
are connected by {\it subdiagrams} consisting of photons and fermion loops,
as shown for example in Fig.~\ref{fig:replicafig}(a). 
Now let us generate $N$ independent copies or {\it replicas} of the gauge and 
fermion fields, labelled by $\{ A_\mu^{(j)} \}$ and $\{ \psi^{(j)} \}$, such that particle 
species with different replica number $j$ never interact. The soft function in such 
a theory, involving the same $n$ Wilson lines (which are not replicated) is 
given by
\beq
  {\cal S}_{n, R} \, = \, \int {\cal D} A^{(1)}_\mu \ldots \int{\cal D}A^{(N)}_\mu
  \exp \left[ {\rm i} e \sum_{j = 1}^N \sum_{i = 1}^n \int dx_i^\mu A^{(j)}_\mu
  + \sum_{j = 1}^N S \left( A_\mu^{(j)}, \bar{\psi}^{(j)}, \psi^{(j)} \right) \right] \, .
\label{replicatheory}
\eeq
Note that the sum in the Wilson line term in \eq{replicatheory} is over both the 
replica numbers and the external lines, since all replicated gauge fields may
interact with any given Wilson line. Furthermore, the fact that the action for the 
replicated theory is just the sum of the actions of individual replicas follows from 
the fact that replicas are non-interacting. Carrying out the path integral in the 
replicated theory amounts to generating Feynman diagrams such as that shown 
in Fig.~\ref{fig:replicafig}(b). Any such diagram must be built of connected 
subdiagrams, such as $G$ and $H$ in the figure, and each individual connected 
subdiagram must contain only a {\it single} replica number, given that the replicated 
gauge fields only interact with their respective replicated fermions, and with the 
Wilson lines. 

The replicated soft function in \eq{pathintegral} is therefore related to the original 
soft function simply by
\beq
   {\cal S}_{n, R} \, = \, {\cal S}_n^N \, ,
\label{SRN}
\eeq
which can be expanded in powers of $N$ to obtain
\beq
  {\cal S}_{n, R} \, = \, 1 + N \log \left( {\cal S}_n \right) + {\cal O} (N^2) \, .
\label{SRN2}
\eeq
It follows that one may write
\beq
  {\cal S}_n \, = \, \exp \left[ \sum _W W \right] \, ,
\label{Sexp}
\eeq
where the sum is over diagrams $W$ that are precisely ${\cal O}(N)$ in the
replicated theory. To find these, note that mutual independence of the replicated 
fields implies that a diagram containing $m$ connected subdiagrams must be
${\cal O} (N^m)$, given that there is a choice of $N$ possible replicas for each 
subdiagram. Thus, the logarithm of the soft function in QED must contain only 
{\it connected subdiagrams}. This result was originally derived using detailed 
combinatorial arguments~\cite{Yennie:1961ad}, which are rather elegantly
circumvented using the replica approach.

In QCD, the combinatorics of exponentiation becomes more complicated due 
to the non-commuting nature of the emission vertices coupling gluons to the 
Wilson lines. Nevertheless, the replica trick argument still works~\cite{Laenen:2008gt}, 
and leads to conclude that the logarithm of the soft function, for processes involving 
only two partons, is built with subdiagrams that are two-line irreducible, which 
were dubbed {\it webs} in the pioneering work of Refs.~\cite{Gatheral:1983cz,
Frenkel:1984pz,Sterman:1981jc}. Similar methods apply to the case of three 
partons, but when more than three coloured particles are involved the nature 
of webs becomes more complicated, due to the multiple possible colour flows 
contributing to the amplitude. Again, however, the replica trick can be used 
to reconstruct  the logarithm of the soft function~\cite{Gardi:2010rn}. In the 
multi-parton case, webs turn out to be sets of diagrams related to each other 
by permutations of gluon attachments to the Wilson lines~\cite{Gardi:2010rn,
Gardi:2011yz}. Multi-parton webs are governed by interesting mathematical 
objects known as {\it web mixing matrices}, whose combinatorial properties 
are continuing to be explored~\cite{Dukes:2013wa,Dukes:2013gea,Dukes:2016ger}.

The arguments just discussed apply directly only to the case of virtual
contributions to the soft function, which arise from a single vacuum expectation 
value of Wilson lines. Including also real emissions, we must define the soft 
function according to \eq{Sdef}, which contains two expectation values involving
non-trivial external states, as well as integrals over the multi-gluon phase space. 
This does not prevent us from using the replica trick: the arguments of this appendix 
can be used to straightforwardly prove exponentiation at cross-section level, 
provided real radiations associated with different replica numbers are mutually 
independent. The latter requirement is fulfilled if the phase space integral for 
$n$ gluon emissions factorises into $n$ decoupled single-gluon phase space
integrals. This condition is satisfied at LL accuracy, as discussed in \secn{sec:LP}.

In this brief summary, we have explicitly discussed only leading-power soft
effects, such that the soft function is defined in terms of vacuum expectation 
values of conventional Wilson lines, as in \eq{Sdef}. The argument, however, 
readily generalises to the next-to-soft function defined in \eq{Stildedef}, which 
involves the generalised Wilson lines of \eq{fidef}. Crucial in the definition of 
\eq{Stildedef} is that the sum over final states involves only leading-power
(and therefore uncorrelated) phase space integrals for $n$ gluon emissions. 
Thus, the replica trick is not invalidated, given that emissions of different gluon 
replicas remain independent, even at next-to-soft level. 


\section{Mellin transforms of NLP contributions}
\label{app:Mellin}

In this Appendix, we collect known results concerning the Mellin transforms
of logarithmic threshold contributions to hadronic cross sections, both at
leading and next-to-leading power. The relevant integrals that need to be 
performed in order to compute the Sudakov exponent at LP and NLP can
be written as
\beq
  {\cal D}_p (N) = \int_0^1 dz \, \frac{z^{N - 1} - 1}{1 - z} \, \ln^p (1 - z) 
  \, , \qquad
  {\cal J}_p (N) = \int_0^1 dz \, z^{N - 1} \, \ln^p(1-z) \, .
\label{intdef}
\eeq
These integrals were computed to the required accuracy (that is, up 
to corrections suppressed by $N^{-2}$ at large $N$) for example in 
Ref.~\cite{Laenen:2008ux}, with the results
\beqa
  {\cal D}_p (N) & = & \frac{1}{p + 1} \, \sum_{k = 0}^{p + 1} 
  d_k (N) \, \binom{p + 1}{k} \, (- \ln N)^{p + 1 - k}
  + {\cal O} \left( \frac{\ln^m N}{N^2} \right) \, , \nn \\
  {\cal J}_p (N) & = & \frac{1}{N} \, \sum_{k = 0}^p 
  \Gamma^{(k)} (1) \, \binom{p}{k} \, ( - \ln N)^{p - k}
  + {\cal O} \left( \frac{\ln^m N}{N^2} \right) \, ,
\label{finDJ}  
\eeqa
where $\Gamma^{(k)}$ is the $k$-th derivative of the $\Gamma$ function,
while
\beq
  d_k (N) \, \equiv \, \frac{d^k}{d \lambda^k} \left[ \Gamma(1 +
  \lambda) \left(1 + \frac{\lambda(1 - \lambda)}{2N} \right) 
  \right]_{\lambda = 0} \, .
\label{defgamk} 
\eeq
Keeping only leading logarithms at both LP and NLP, one finds
\beq
  {\cal D}_p (N) \, = \, (-1)^{p + 1} \left[ \, \frac{1}{p + 1} \log^{p + 1} N -
  \frac{\log^p N}{2 N} \, \right] + \ldots \, ,
\label{Mplus2}
\eeq
as well as
\beq
  {\cal J}_p (N) \, = \, \frac{ \left( - \log N \right)^p}{N} + \ldots \, .
\label{Mlog2}
\eeq
Considering now the application of these results to \eq{NEresum}, we note that
the partonic factor for the ($q \bar{q}$)-channel of the resummed Drell-Yan 
cross-section at LL accuracy, at ${\cal O}(\alpha_s^m)$, and in Mellin space, 
takes the form
\beqa
  && \hspace{-1cm} \left( \frac{2 \alpha_s C_F}{\pi} \right)^m \frac{1}{m!}
  \left( \log^2 N + \frac{\log N}{N} \right)^m \nn \\
  & = & \left( \frac{2 \alpha_s C_F}{\pi} \right)^m \frac{1}{(m-1)!}
  \left[ 2 \left( \frac{\log^{2m} N}{2m} - \frac{\log^{2 m - 1} N}{2 N} \right)
  + \frac{2 \log^{2 m - 1} N}{N} \right] \, .
\label{invmel}
\eeqa
In the second line, we have rewritten the result in order to explicitly recognise 
the leading-logarithmic contributions to the integrals ${\cal D}_{2 m  - 1}$ and 
${\cal  J}_{2 m  - 1}$, given in \eq{Mplus2} and in \eq{Mlog2}, respectively. One
finds then
\beq
  \left( \frac{2 \alpha_s C_F}{\pi} \right)^m \frac{2}{(m - 1)!}
  \Big( {\cal D}_{2 m - 1} (N) - {\cal J}_{2 m - 1} (N) \Big) \, ,
\eeq
which leads immediately to \eq{LLNLP2}.


\section{Two gluon emission from the generalised Wilson line}
\label{app:twogluon}

In \secn{sec:NLPmatrix}, we defined a {\it next-to-soft} function in terms of 
generalised Wilson lines, which have been introduced and discussed extensively 
in Refs.~\cite{Laenen:2008gt,Laenen:2010uz}. These operators generate
effective Feynman rules for the emission of (next-to-)soft gluons from
a given hard particle, and the one-gluon emission terms required for
describing radiation at ${\cal O}(\alpha_s)$ are shown in \eq{fidef}. However, 
as Refs.~\cite{Laenen:2008gt,Laenen:2010uz} make clear, the required Feynman 
rules also involve effective vertices describing the emission of two gluons from 
the same point. These are neglected in the analysis of this paper, for reasons 
discussed in \secn{sec:NLPmatrix}. It is therefore appropriate to check explicitly
in a simple example that such vertices cannot contribute to leading-logarithmic 
NLP terms at higher orders in perturbation theory. 

Ignoring coupling and colour factors, the form of the two-gluon emission vertex 
from a hard scalar particle, in momentum space, is given by~\cite{Laenen:2008gt}
\beq
  R_{\mu \nu} (p;k,l) \propto \frac{(p \cdot k)(p \cdot l) \eta^{\mu\nu} -
   p^\nu l^\mu (p \cdot k) - p^\mu k^\nu (p \cdot l)
  +  (k \cdot l) p^\mu p^\nu}{(p \cdot k) (p \cdot l) \left[ p \cdot (k + l) \right]} \, ,
\label{Rmunu}
\eeq
where $p$ is the hard momentum of the emitting particle, and $(k, l)$ are the soft
momenta of the emitted gluons. The latter may also be sums of individual
gluon momenta, which will not affect the following. 

Throughout the paper, we have considered processes with two incoming
massless hard partons carrying four-momenta $p_1$ and $p_2$. Without 
loss of generality, let us consider the two-gluon emission vertex as occuring 
on leg $p_1$. Then, as we have argued in \secn{sec:NLPmatrix}, leading
logarithmic effects can only come from radiation that is maximally (next-to) soft, 
as well as collinear. This in turn means that either $k$ or $l$ must be proportional 
to $p_1^\mu$ or $p_2^\mu$. From \eq{Rmunu}, it is straightforward to show 
that $R(p; k, l)$ vanishes if $k \propto p$ or if $l \propto p$. Thus, for a non-zero
contribution, both $k$ and $l$ must be proportional to $p_2$, yielding
\beq
  R^{\mu\nu} (p_1; p_2, p_2) \, = \, \frac{1}{4 (p_1 \cdot p_2)}
  \left[ \eta^{\mu\nu} - \frac{ \left( p_1^\mu p_2^\nu + p_1^\nu p_2^\mu 
  \right)}{p_1 \cdot p_2} \right] \, .
\label{Rmunu2}
\eeq
In a squared matrix element summed over final state gluon polarisations, the 
Lorentz indices $\mu$ and $\nu$ must ultimately be contracted with one of the 
external momenta $p_1$ or $p_2$, or with a further soft momentum. However, 
the combination in the square brackets in \eq{Rmunu2} acts as a projection 
tensor, that removes the component of any four-momentum that is collinear 
with $p_1$ or $p_2$. We have already seen that leading log behaviour can 
only arise from soft gluon emissions that are maximally (next-to) soft and
collinear. We thus find that the two-gluon emission vertex is irrelevant at LL
accuracy.

To be more precise, the above discussion relates only to emissions from a 
scalar particle. In the case of non-zero spin, an extra contribution to the
two-gluon emission vertex appears, that involves the spin generator of the 
emitting particle. There is however an independent line of argument that 
allows us to discard this contribution, and that indeed could be applied to 
the first term of \eq{Rmunu}: in position space, a four-point vertex for double 
gluon emission necessarily involves both gluons being emitted from the same 
point on the emitting Wilson line, and thus involves one less propagator 
than contributions involving two separate gluon emissions. As a result, 
such contributions will not contribute a leading logarithm which, as 
discussed in \secn{sec:NLPmatrix}, requires a maximal number of
integrations over normal variables. We therefore conclude, also in the
case of spinning hard particles, that the two-gluon next-to-soft
emission vertex can be neglected at LL accuracy.


\bibliography{refs}


\end{document}